\documentclass{spie}
\usepackage[utf8]{inputenc}

\usepackage{amsmath,amsfonts,amssymb}
\usepackage{graphicx}
\usepackage[colorlinks=true, allcolors=blue]{hyperref}
\usepackage{appendix}

\usepackage{aas_macros}

\usepackage{scalerel,stackengine}
\stackMath
\newcommand\reallywidehat[1]{%
\savestack{\tmpbox}{\stretchto{%
  \scaleto{%
    \scalerel*[\widthof{\ensuremath{#1}}]{\kern-.6pt\bigwedge\kern-.6pt}%
    {\rule[-\textheight/2]{1ex}{\textheight}}
  }{\textheight}%
}{0.5ex}}%
\stackon[1pt]{#1}{\tmpbox}%
}
\title{Pair-based Analytical model for Segmented Telescopes Imaging from Space (PASTIS) for sensitivity analysis}

\author{Lucie Leboulleux\supit{a}\supit{b}\supit{c}, Jean-Fran\c{c}ois Sauvage\supit{b}\supit{c}, Laurent Pueyo\supit{a}, Thierry Fusco\supit{b}\supit{c}, R\'{e}mi Soummer\supit{a}, Johan Mazoyer\supit{d}, Anand Sivaramakrishnan\supit{a}\supit{d}, Mamadou N'Diaye\supit{e}, Olivier Fauvarque\supit{b}
\skiplinehalf
\supit{a} Space Telescope Science Institute, 3700 San Martin Drive, Baltimore, MD 21218, USA\\
\supit{b} Aix Marseille Universit\'{e}, CNRS, LAM (Laboratoire d'Astrophysique de Marseille) UMR 7326, 13388, Marseille, France\\
\supit{c} Office National d'Etudes et de Recherches A\'{e}rospatiales, 29 Avenue de la Division Leclerc, 92320 Ch\^{a}tillon, France\\
\supit{d} Department of Physics and Astronomy, Johns Hopkins University, Baltimore, MD, USA \\
\supit{e} Universit\'{e} C\^{o}te d\textquoteright Azur, Observatoire de la C\^{o}te d\textquoteright Azur, CNRS, Laboratoire Lagrange, Bd de l\textquoteright Observatoire, CS 34229, 06304 Nice Cedex 4, France
}
\authorinfo{Further author information, send correspondence to Lucie Leboulleux: E-mail: leboulleux@stsci.edu, Telephone: 1 410 338 2881}
\begin{document}
\maketitle

\begin{abstract}

The imaging and spectroscopy of habitable worlds will require large-aperture space-based telescopes, to increase the collecting area and the angular resolution. These large telescopes will necessarily use segmented primaries to fit in a rocket. However, these massively segmented mirrors make high-contrast performance very difficult to achieve and stabilize, compared to more common monolithic primaries. Despite space telescopes operating in a friendlier environment than ground-based telescopes, remaining vibrations and resonant modes on the segments can still deteriorate the performance. 

In this context, we present the Pair-based Analytical model for Segmented Telescopes Imaging from Space (PASTIS) that enables the establishment of a comprehensive error budget, both in term of segment alignment and stability. Using this model, one may evaluate the influence of the segment cophasing and surface quality evolution on the final images and contrasts, and set up requirements for any given mission. One can also identify the dominant modes of a given geometry for a given coronagraphic instrument and design the feedback control systems accordingly.

In this paper, we first develop and validate this analytical model by comparing its outputs to the images and contrasts predicted by an end-to-end simulation. We show that the contrasts predicted using PASTIS are accurate enough compared to the end-to-end propagation results, at the exo-Earth detection level. Second, we develop a method for a fast and efficient error budget in term of segment manufacturing and alignment that takes into account the disparities of the segment effects on the final performance. This technique is then applied on a specific aperture to provide static and quasi-static requirements on each segment for local piston and $45^\circ$-astigmatism aberrations. Finally we discuss potential application of this new technique to future missions.

\end{abstract}


\section{Introduction}
\label{sec:Intro}

Direct imaging and spectroscopy of Earth-like exoplanets will require future telescopes to be larger. Indeed the science yield increases as a steep power of primary mirror diameter, especially so when using a coronagraph\cite{Stark2016}. In order to fit these mirrors in launch vehicles, these large primary mirrors will have to be segmented. Coronagraphs adapted to these new pupil geometries have already been designed and validated on ground-based telescopes such as the Keck telescopes. Even if coronagraphs on the Keck telescopes could be made to work at smaller inner working angle (IWA), they are dedicated to infrared observations and are limited by the atmosphere, and can therefore only reach a modest contrast\cite{Ruane2017, Mawet2017,Mawet2016}. Space segmented telescopes with similar contrast to Keck are imminent \cite{Boccaletti2015, Krist2007}. Until recently segmented pupil coronagraph designs with sufficient performance to image Earth-like planets did not exist. However the latest developments in coronagraph design promise contrasts of the order of $10^{-10}$\cite{Carlotti2011, Mazoyer2017, Zimmerman2016, Pueyo2013, Guyon2014}. The most recent progress in coronagraphy on monolithic apertures with secondary mirrors and other necessary obstructions is being applied on the Wide Field Infrared Survey Telescope (WFIRST).  This application will demonstrate wavefront sensing and control in the presence of thermal drifts\cite{Krist2016, Shi2016}. However WFIRST does not address stability issues associated with segmentation. For this reason, we need a good understanding of the impact of segment level errors on coronagraphic Point Spread Function (PSF) quality. On the upcoming segmented James Webb Space Telescope (JWST), relevant mission requirements only concern the encircled energy and Strehl ratio\cite{Lightsey2010,Lightsey2014}. In this paper we generalize the error budgeting on contrast requirements with a general tool that is applicable to any segmented pupil.  Our work is also directly applicable to Extremely Large Telescopes (ELTs)\cite{Macintosh2006,Kasper2008,Davies2010,Quanz2015}, albeit at more modest contrasts. In particular, an analytical study has also been driven by Yaitskova et al. \cite{Yaitskova2003} for ELT-like configurations.

Several experiments in high-contrast imaging have produced very encouraging results. The best contrast achieved to date is a few $10^{-9}$.  This was obtained on the High-Contrast Imaging Testbed (HCIT), with a circular aperture in extremely well-controlled conditions\cite{Trauger2007}. A contrast of a few $10^{-8}$ was also reached on the Tr\`{e}s Haute Dynamique (THD) bench, at separations below 0.5 arcsec\cite{Baudoz2012, Mazoyer2014}.  The latter would allow the detection of mature exo-Jupiters. However these experiments do not include segmentation, and moreover, are mostly static. More work is needed to extrapolate these results to our desired contrast. Similar experiments on segmented apertures are needed in order to build future telescopes for exoplanet imaging (such as the Large Ultra-Violet Optical Infrared (LUVOIR) telescope \cite{Dalcanton2015,Pueyo2017} or the Habitable Exoplanet Imaging Mission (HabEx) \cite{Mennesson2016}).

To get stable imaging and maintain sufficient contrast over long times, error budgets must be an integral part of the optical systems being considered. Since numerous factors can degrade the performance of the system, and since the objective is extremely challenging, a comprehensive error budget is essential in order to make wise decisions early enough during developments. Current methods for error budget are simulations of end-to-end propagation through the optical system, with variations of factors that are known to impact the contrast: segment phasing errors, segment surface quality, local or global vibrations, resonant modes on the segments, quasi-static aberrations due to thermal drift, and so on\cite{Stahl2013, Stahl2015}.  Because of the large number of factors that affect contrast, a multitude of cases need to be tested.  Because of the computational burden involved, these studies can be dauntingly slow. Standard error budget methodology relies on multiple random realizations of disturbances, measuring science metrics based on simulated propagation of disturbances and establishing confidence intervals for acceptable operating points, given stated science requirements. In fact even simple metrics such as encircled energy, can be beyond the capabilities of numerical optical propagation.

This is the motivation behind our alternative fast method, which is based on the contrast criterion, and adaptable to any segmented pupil (such as JWST \cite{Greenhouse2016,Clampin2008}, ELTs \cite{Macintosh2006,Kasper2008,Davies2010,Quanz2015}, the HabEx mission \cite{Mennesson2016}, or the LUVOIR telescope \cite{Dalcanton2015,Pueyo2017}). This new method is based on a so-called Pair-based Analytical model for Segmented Telescopes Imaging from Space (PASTIS), an analytical model to directly express the focal plane image and its contrast as a function of the Zernike coefficients of the segments' wavefront aberrations. A simple inversion of the model immediately provides the constraints in cophasing and stability that are necessary for obtaining the desired contrast. In this paper we focus on the development of the analytical model, its validation, its formal inversion, and its application to tolerancing and stability constraints.

In section 2, we introduce our analytical model, which is based on a segment-based model of the pupil with a perfect coronagraph, to enable sufficiently high-contrast performance. In particular, we develop a matrix-based version of the analytical model, which shortens the integrated contrast computation by a factor of the order of $10^7$. In section 3, we apply it to an example of a segmented pupil that we will use for the rest of the paper.  We compare our model output to images created by an end-to-end simulation, where the segmented pupil is combined with an Apodized Lyot Coronagraph (APLC) that enables a $10^{-10}$ contrast in a circular dark hole from $4 \lambda /D$ to $9 \lambda /D$ with a monochromatic light at $\lambda = 640$ nm. In the last section we use this matrix-based analytical model to provide a new method for a tolerancing and stability study on the segment alignment and manufacturing for all segmented pupils by sidestepping the iterative process of traditional error budgeting. Here we apply our method to the cases of local pistons and $45^\circ$-astigmatisms on the segments to provides very results that agree very well with the much slower full optical propagation calculations.

\section{Analytical model of image formation and contrast with a segmented pupil through a coronagraph}
\label{sec:AM}

In section~\ref{sec:Image formation with phase aberrations} we use the development of Bordé and Traub\cite{Borde2006} to express the image in the final plane as a function of the aberrations in the pupil, behind a perfect coronagraph.

\begin{figure}
   \begin{center}
   \begin{tabular}{cccc}
   \includegraphics[height=3.8cm]{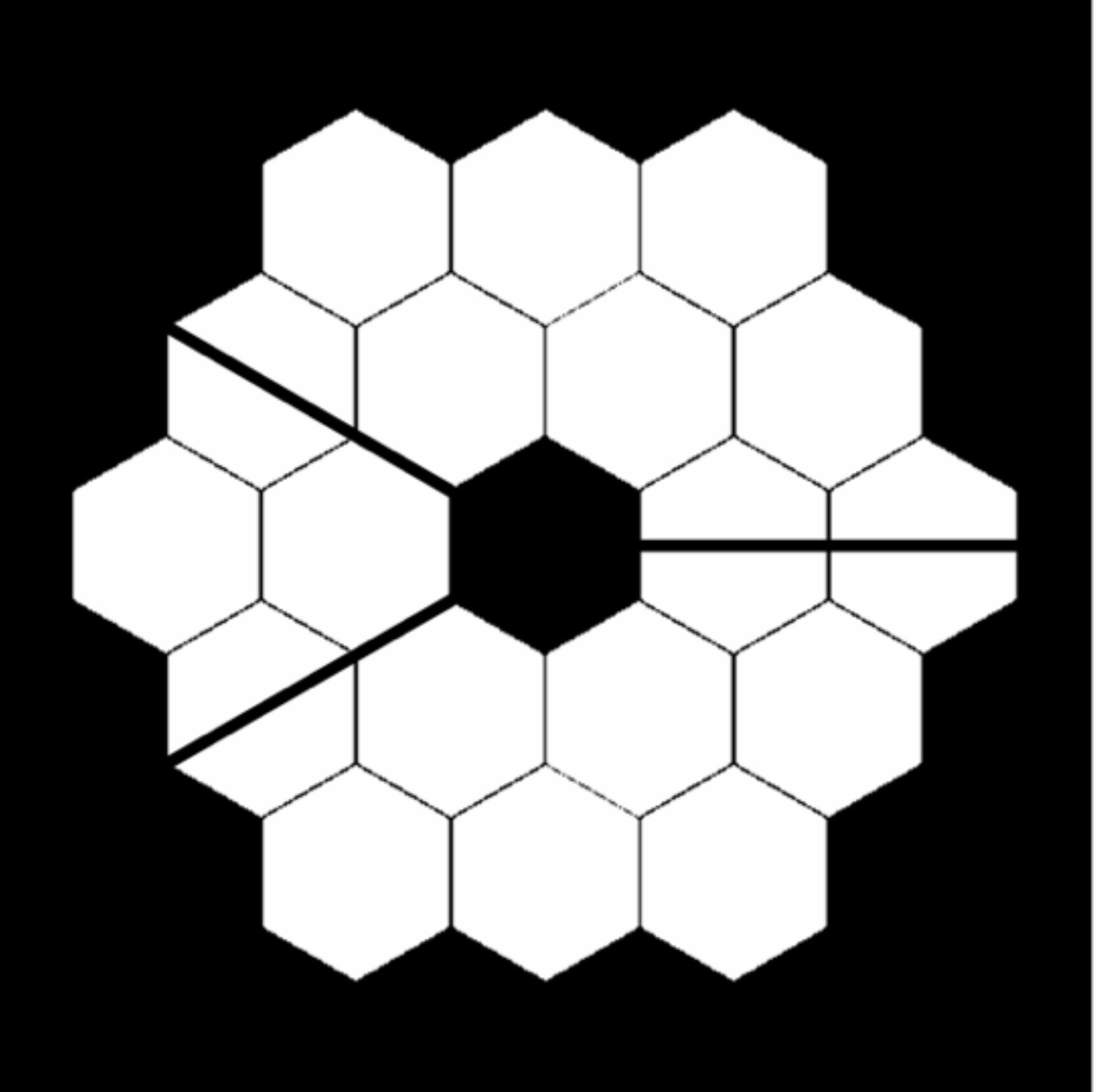} &
    \includegraphics[height=3.8cm]{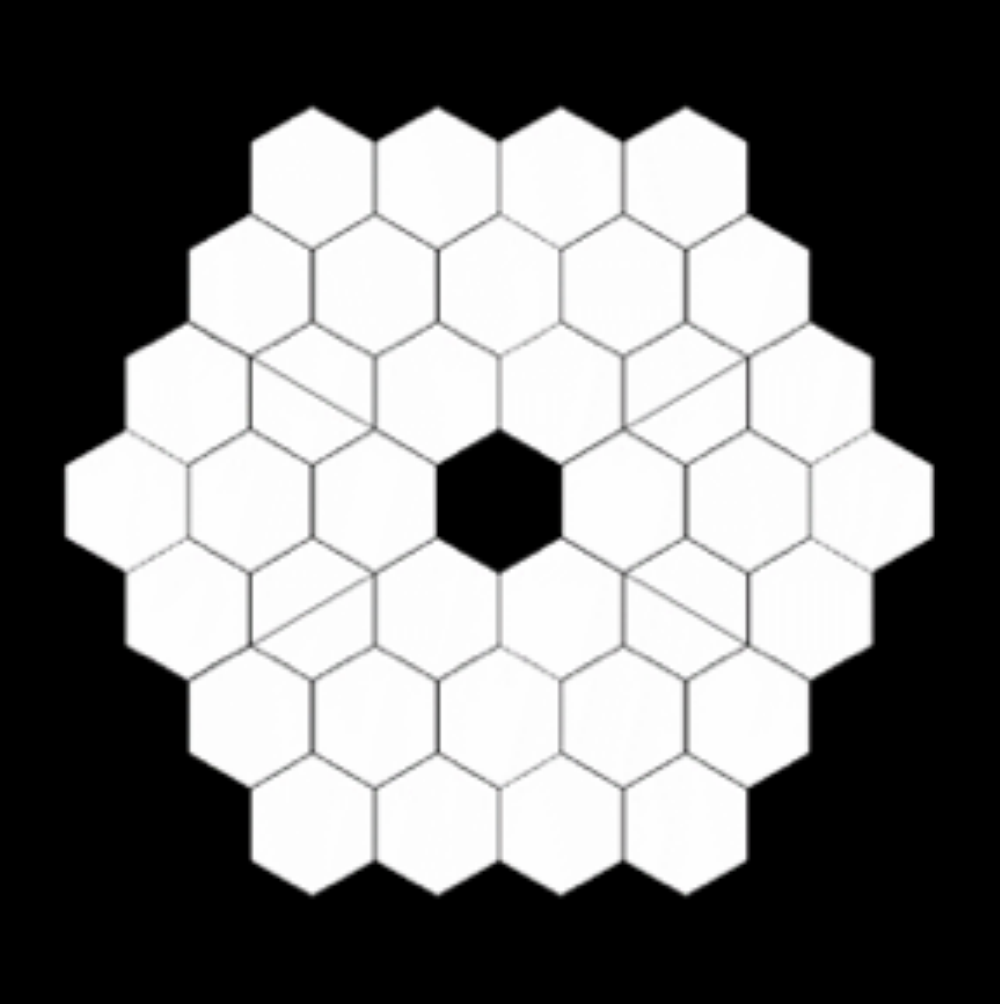} &
    \includegraphics[height=3.8cm]{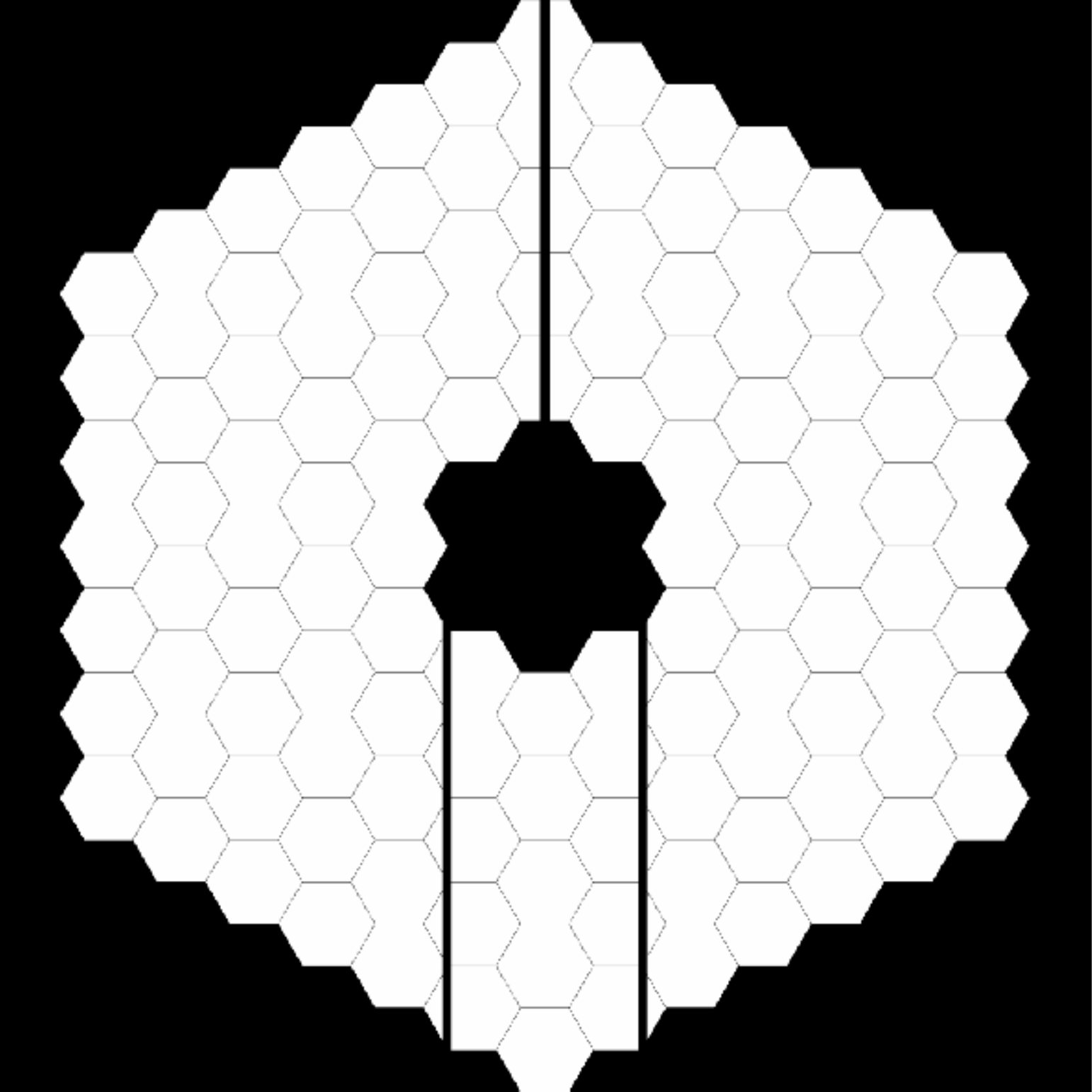} &
   \includegraphics[height=3.8cm]{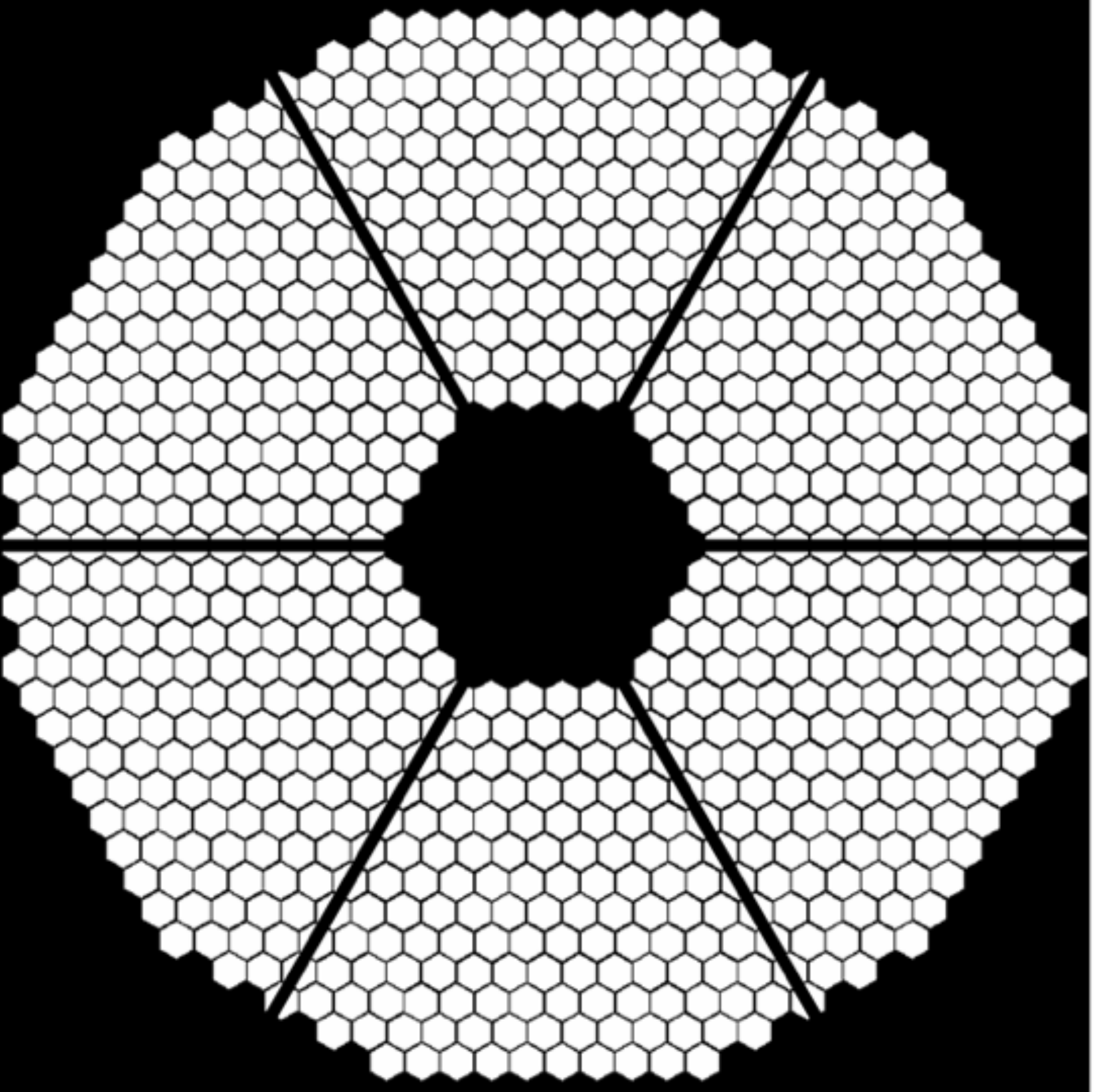}\\
   (a) & (b) & (c) & (d)
   \end{tabular}
   \end{center}
   \caption[Segmented_Pupils] 
   { \label{fig:Segmented_Pupils} 
(a) The JWST pupil (18 segments), (b) a pupil used for the Segmented Coronagraph Design and Analysis (SCDA) study (36 segments), (c) one of the optional LUVOIR pupils (120 segments), and (d) the ELT pupil (798 segments). The pupil (b) will be used for applications later in this paper.}
   \end{figure}

The model then developed in sections~\ref{sec:Case of a segmented pupil} and~\ref{sec:M-AM} is applicable to all segmented pupils composed from the repetition of a generic segment. A few examples are indicated in Fig.~\ref{fig:Segmented_Pupils}.

\subsection{Image formation with phase aberrations}
\label{sec:Image formation with phase aberrations}

The electric field in the pupil plane is
\begin{equation} \label{eq:31}
    E(\mathbf{r}) = P(\mathbf{r})e^{\alpha(\mathbf{r}) + i \phi(\mathbf{r})}.
\end{equation}
where $P$ is the entrance pupil of our optical system, $\alpha$ are the amplitude aberrations, and $\phi$ are the phase aberrations in the pupil. Even if the amplitude errors are an important point in coronagraphy, they are not considered in this analytical model. The analytical formalism presented herein can however be readily generalized to include amplitude aberrations. Such considerations are left for further studies.

Since the phase aberrations are small, equation~\ref{eq:31} leads to
\begin{equation}
    E(\mathbf{r}) = P(\mathbf{r}) +i P(\mathbf{r})\phi(\mathbf{r}).
\end{equation}
Since the phase aberration is defined on the pupil, this expression can be simplified using $P(\mathbf{r})\phi(\mathbf{r}) = \phi(\mathbf{r})$. Thus the amplitude of the electric field in the detector plane becomes
\begin{equation} \label{eq:2}
    E_f(\mathbf{u}) = \widehat{P}(\mathbf{u}) + i \widehat{\phi}(\mathbf{u}),
\end{equation}
where $\mathbf{u}$ is the position vector in the detector plane (focal plane) and $\hat{f}$ is the Fourier Transform of the function $f$.

Without coronagraph, the image intensity is:
\begin{equation}
    \hspace{-5em}I(\mathbf{u}) = \left \Vert E_f(\mathbf{u}) \right \Vert ^2
\end{equation}
\begin{equation} \label{eq:10}
    \hspace{9.5em} = \left \Vert \widehat{P}(\mathbf{u}) \right \Vert ^2 + \left \Vert \widehat{\phi}(\mathbf{u}) \right \Vert ^2 + 2 \Re(\widehat{P}^*(\mathbf{u}) i \widehat{\phi}(\mathbf{u}))
\end{equation}
This equation is made of a constant term, a term linear to the aberrations, and a quadratic term.

However, in the hypothesis of a perfect coronagraph, the amplitude of the on-axis electric field generated by the star is removed \cite{Malbet1995,Quirrenbach2005,Cavarroc2006}. The intensity becomes:
\begin{equation}
    \hspace{2em}I(\mathbf{u}) = \left \Vert E_f(\mathbf{u}) - \widehat{P}(\mathbf{u}) \right \Vert ^2
\end{equation}
\begin{equation} \label{eq:44}
     = \left \Vert \widehat{\phi}(\mathbf{u}) \right \Vert ^2
\end{equation}

\subsection{Case of a segmented pupil}
\label{sec:Case of a segmented pupil}

The pupil is now considered segmented in several identical segments.

\subsubsection{Pupil and phase models}

\begin{figure}
   \begin{center}
   \begin{tabular}{cc}
   \includegraphics[height=6cm]{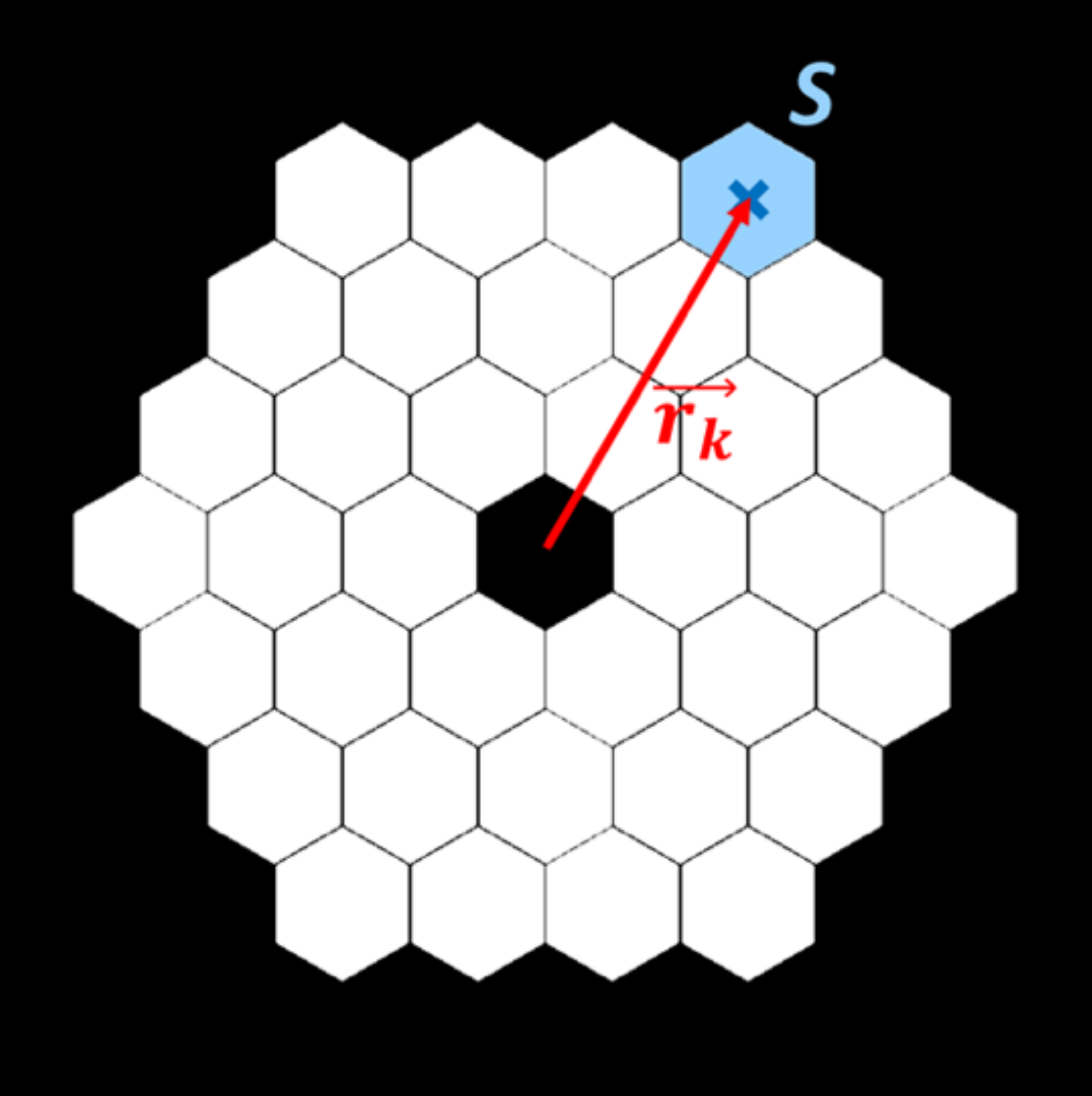} & 
   \includegraphics[height=6cm]{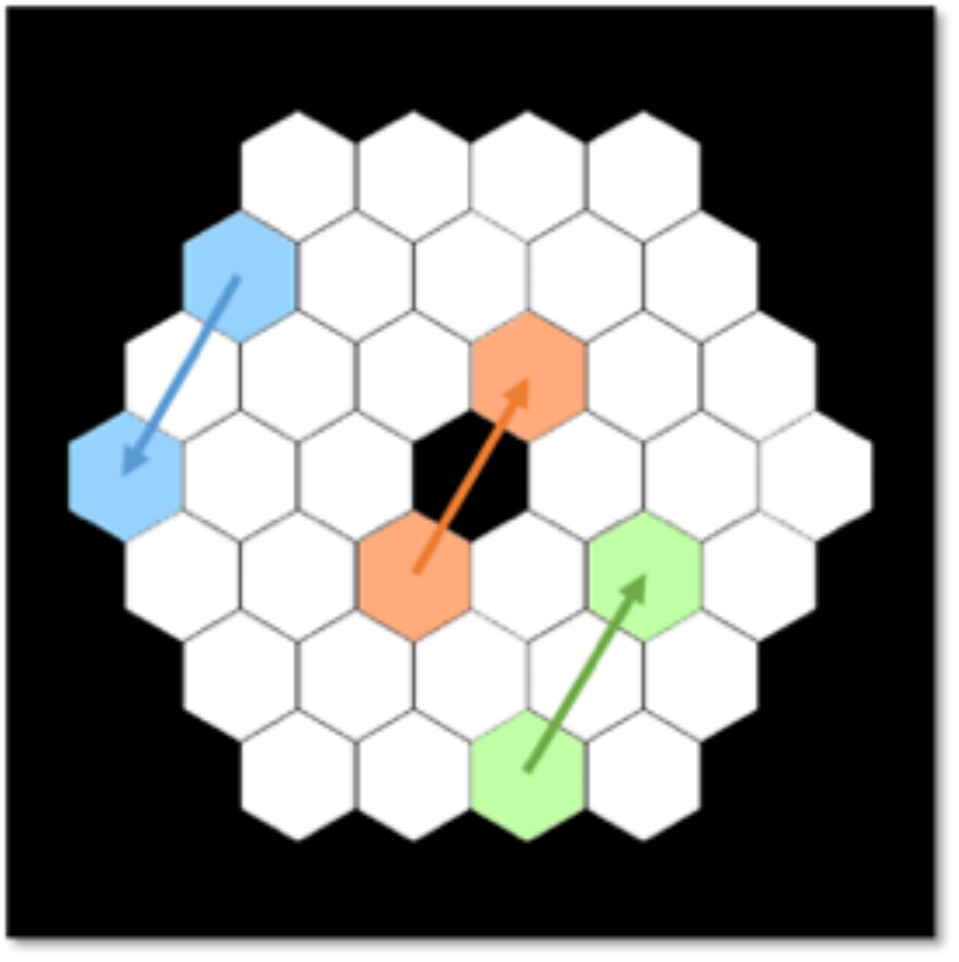} \\
   (a) & (b)
   \end{tabular}
   \end{center}
   \caption[schema_analytic_model] 
   { \label{fig:schema_analytic_model} 
(a) Definitions of the vectors $\mathbf{r_k}$ and of the shape of a segment $S$ on a segmented pupil, here the SCDA primary mirror. In red, we can see one of the vector $\mathbf{r_{k}}$, from the center of the pupil to the $k$-th segment, expressed in pixels. (b) Illustration of some redundant oriented pairs that correspond to one single non-redundant pair. 42 oriented pairs generate exactly the same interference fringes than the pair $\mathbf{r_{16}}-\mathbf{r_{28}}$ (blue), for example the pairs $\mathbf{r_{25}}-\mathbf{r_{12}}$ (orange) and $\mathbf{r_{14}}-\mathbf{r_3}$ (green). Since these 42 pairs have the same effect in the detector plane, they can all be replaced by one single pair, called the non-redundant pair.}
   \end{figure}

The entrance pupil $P$ of our optical system consists of $n_{seg}$ identical segments of shape mask $S$. We define the pupil as
\begin{equation}
    P(\mathbf{r})=\sum_{k=1}^{n_{seg}} S(\mathbf{r}-\mathbf{r_k}),
\end{equation}
where $\mathbf{r}$ is the position vector in the pupil plane, and $\mathbf{r_k}$ the position of the center of the $k$-th segment (Fig.~\ref{fig:schema_analytic_model}).

\begin{figure}
   \begin{center}
   \begin{tabular}{cc}
   \includegraphics[height=5.5cm]{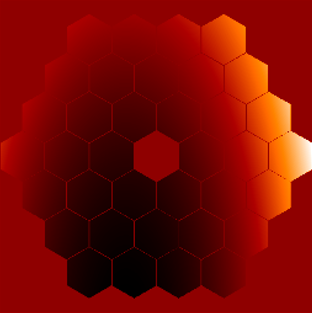} & \includegraphics[height=5.5cm]{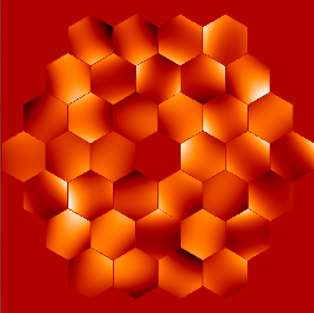}
   \end{tabular}
   \end{center}
   \caption[Global_Local] 
   { \label{fig:Global_Local} 
Left: Global aberrations on a segmented mirror. Right: Local aberrations on the same segmented mirror.}
   \end{figure}

We focus here on phasing, alignment, or polishing errors of the primary mirror. The phase aberration $\phi$ on the pupil $P$ is expressed as the sum of global and local aberrations on the different segments (see Fig.~\ref{fig:Global_Local}). Since the global aberrations can be seen as a sum of local aberrations\cite{Janinpotiron2018}, we simply express the phase in the pupil as:
\begin{equation} \label{eq:1}
    \phi(\mathbf{r})=\sum_{(k,l)=(1,1)}^{(n_{seg},n_{zer})} a_{k,l} Z_l(\mathbf{r}-\mathbf{r_k}),
\end{equation}
where the coefficients $(a_{k,l})_{(k,l)\in \lbrack 1,n_{seg} \rbrack \ times \lbrack 1,n_{zer} \rbrack}$ are the local Zernike coefficients and $(Z_l)_{l\in \lbrack 1,n_{zer} \rbrack}$ is the Zernike basis on a circular pupil that overfills the segment cropped by the shape of the generic segment. We could here have used the basis of polynomials specific to hexagonal apertures\cite{Mahajan2006, janinpotiron2017}. However since this basis' vectors are linear combinations of the common circular-aperture Zernike polynomials, both studies are quite equivalent, in particular for low-order polynomials and we decided to keep using the well-known Zernike polynomials basis.

Manufacturing, telescope alignment, and initial primary mirror cophasing remove most of the global and local aberrations, leaving only residual aberrations. Further studies could be done in the case of misalignment of the secondary mirror, which is known to impact significantly the performance of the system. However, in this article, the model is developed in the case of residual errors on the primary mirror only. 

\subsubsection{Case of one single Zernike on the segments}
\label{sec:Case of one single Zernike on the segments}

We first study the case where only one local Zernike polynomial $Z_l$ is applied on the segments, even if the coefficient $a_{k,l}$ can vary with the segment $k$ (see Fig.~\ref{fig:PSF} for some examples of phases). 

The intensity with a perfect coronagraph derives from equations~\ref{eq:44} and ~\ref{eq:1}, this second one being used in the case of one single Zernike polynomial:
\begin{equation} 
    \hspace{-3em} I(\mathbf{u}) = \left \Vert \widehat{Z_l}(\mathbf{u}) \sum_{k=1}^{n_{seg}} a_{k,l} e^{-i \mathbf{r_k} . \mathbf{u}} \right \Vert ^2
\end{equation}
\begin{equation} \label{eq:10000}
    \hspace{3em} = \left \Vert \widehat{Z_l}(\mathbf{u}) \right \Vert ^2 \sum_{i=1}^{n_{seg}} \sum_{j=1}^{n_{seg}} a_{i,l} a_{j,l} e^{i (\mathbf{r_{j}} - \mathbf{r_{i}} ). \mathbf{u}}
\end{equation}
Since the intensity is real, $\sum_{i=1}^{n_{seg}} \sum_{j=1}^{n_{seg}} a_{i,l} a_{j,l} e^{i (\mathbf{r_{j}} - \mathbf{r_{i}} ). \mathbf{u}}$ is real, and therefore:
\begin{equation}\label{eq:67}
    I(\mathbf{u}) = \left \Vert \widehat{Z_l}(\mathbf{u}) \right \Vert ^2 \bigg( \sum_{k=1}^{n_{seg}} a_{k,l} ^2 + \sum_{i=1}^{n_{seg}} \sum_{j=1,j \ne i}^{n_{seg}} a_{i,l} a_{j,l} \cos((\mathbf{r_{j}} - \mathbf{r_{i}} ). \mathbf{u})\bigg)
\end{equation}

It appears here that studying the effect of random values of the same Zernike on all the segments is equivalent to studying the interference effects on each pair of segments and summing them.

$\bigg( \mathbf{u} \rightarrow \cos((\mathbf{r_{j}} - \mathbf{r_{i}} ). \mathbf{u})\bigg)_{(i,j)\in \lbrack 1,n_{seg} \rbrack^2}$ is a basis of cosine functions that describes the intensity distribution in the focal plane. Each vector of this basis is a cosine function whose spatial periodicity and orientation depends on the pair of considered segments $(i,j)$, very similarly to the Young fringes\cite{BornWolf1999}. The amplitude in front of this function is simply the product of the amplitudes $a_{i,l}$ and $a_{j,l}$ of the respective Zernike coefficients on each segment. As a consequence, two pairs of segments having an identical configuration result in the same intensity pattern (see Fig.~\ref{fig:schema_analytic_model}).

\begin{figure}
   \begin{center}
   \begin{tabular}{c}
   \includegraphics[height=7cm]{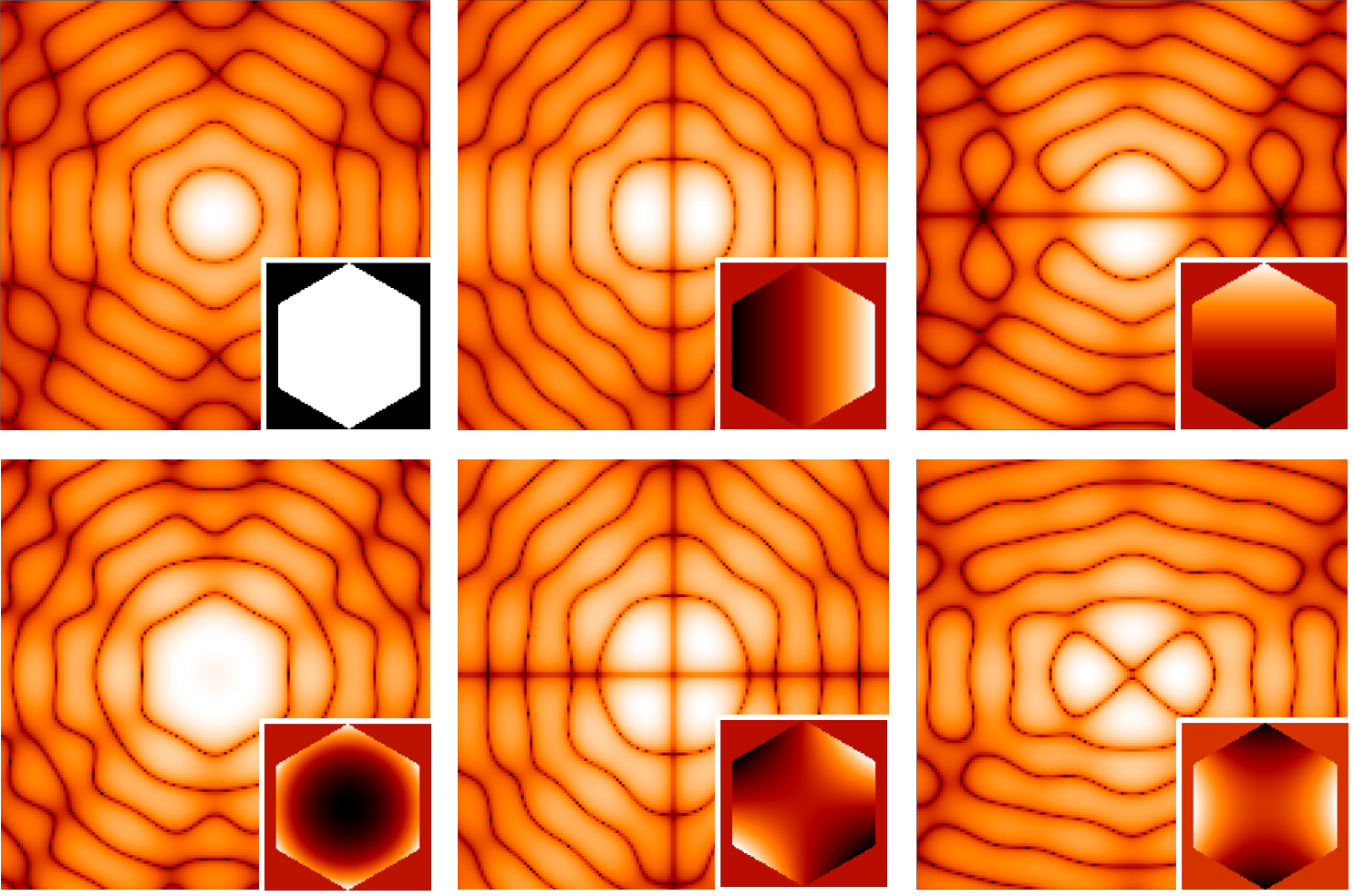}
   \end{tabular}
   \end{center}
   \caption[enveloppes] 
   { \label{fig:enveloppes} 
Envelopes corresponding to the first few Zernike polynomials, in logarithmic scale of the intensity, from 0 to $35\lambda/D$. Top left: piston, top center: tip, top right: tilt, bottom left: focus, bottom center: $45^{\circ}$-astigmatism, bottom right: $0^{\circ}$-astigmatism.}
   \end{figure}

This equation shows also that the local Zernike polynomial $Z_l$ acts on the final image plane as an envelope, which does not depend on the segment positions. Fig.~\ref{fig:enveloppes} illustrates the envelopes for the first Zernike polynomials.

We call $n_{NRP}$ the number of non-redundant segment pairs and $(\mathbf{b_q})_{q\in \lbrack 1,n_{NRP} \rbrack}$ the basis of non-redundant segment pairs. This basis corresponds to the family of all the vectors joining the centers of two different segments, where each vector appears once. Fig.~\ref{fig:schema_analytic_model} (b) illustrates the redundancy of some pairs of segments: the three vectors represented here are represented by a unique vector in the basis of non-redundant segment pairs. In the case of the SCDA pupil, which contains 36 segments, there are 1260 possible oriented pairs of segments $\mathbf{r_{j}} - \mathbf{r_{i}}$ (obtained with the binomial coefficient $2 \times C^2_{36}$), but $n_{NRP}=63$. In the case of JWST, there are 18 segments, 306 pairs of segments in total, but only 30 non-redundant pairs of segments. In the case of the ELT, there are 798 segments, 636006 pairs of segments in total, and 1677 non-redundant pairs of segments. In all these examples, just a small fraction of the segment pairs, the non-redundant vectors, is responsible of the pattern in the focal plane\cite{Tuthill2000, Lacour2011}.

Thanks to this remark, we can rearrange the double sum of equation~\ref{eq:67} as
\begin{equation}
\label{eq:formule}
    I(\mathbf{u}) = \left \Vert \widehat{Z_l}(\mathbf{u}) \right \Vert ^2 (\sum_{k=1}^{n_{seg}} a_{k,l} ^2 + 2 \sum_{q=1}^{n_{NRP}} A_q \cos(\mathbf{b_q} . \mathbf{u})),
\end{equation}
with
\begin{equation}
    \forall q\in \lbrack 1,n_{NRP} \rbrack, A_q = \sum_{(i,j)} a_{i,l} a_{j,l}
\end{equation}
where the couples $(i,j)$ are all the couples that verify the relation $\mathbf{r_{j}}-\mathbf{r_{i}}=\pm \mathbf{b_q}$. This way the matrix multiplications associated with the focal plane sample arrays only has to be done $n_{NRP}$ times instead of $n_{seg}^2$ times. This new equation saves computing time and resources and is strictly and mathematically equivalent to the equation~\ref{eq:67}: instead of summing independently all the cosines with their coefficients (equation~\ref{eq:67}), we unify all the identical cosines and sum up their amplitudes (equation~\ref{eq:formule}). 

We can conclude that it is possible to obtain a relation between the final image, a certain baseline, and the Zernike coefficients applied on each segment of the baseline.

\subsubsection{Case of an actual coronagraph}
\label{sec:Calibration}

Equation~\ref{eq:formule} establishes that the intensity in the image can be expressed as a function of the Zernike coefficients applied on the pupil. In this formula, all the individual segments have the same contribution to the final image pattern and its contrast. An actual coronagraphic system can include pupil apodizer, pupil phase, focal plane mask (phase or amplitude), and a Lyot stop. For the sake of our example in this paper, we only study pupil apodization, hard edge focal plane mask, and hard edge Lyot stop. These optical components, added to the spiders of the telescope, introduce a dependency on the segments: all the segments are not strictly equivalent. Therefore, the model needs to be refined.

In the case of direct imaging with a perfect coronagraph, we ignored the direct PSF $\widehat{P}$ in the expression of the electric field in the final plane $E_f$. A real coronagraph has different impacts on the image. The coronagraph function, modeled here as a linear function $\mathcal{C}$. In this general case, the intensity becomes:
\begin{equation}
    \hspace{-6.5em} I(\mathbf{u}) = \left \Vert \mathcal{C}\{P\} + i\mathcal{C}\{\phi\} \right \Vert ^2 
\end{equation}
\begin{equation} \label{eq:100}
    \hspace{5em} = \left \Vert \mathcal{C}\{P\} \right \Vert ^2 + 2 \Re{\{\mathcal{C}\{P\}\mathcal{C}\{\phi\}^*\}} + \left \Vert \mathcal{C}\{\phi\} \right \Vert ^2
\end{equation}
While it might seem different than Eq.~\ref{eq:44}, it can still be very close:

- if $\phi$ is large enough, then $\left \Vert \mathcal{C}\{\phi\} \right \Vert ^2$ dominates over the rest of the equation. Since $\mathcal{C}\{\phi\} = \sum_{k=1}^{n_{seg}} \mathcal{C}\{a_{k,l} Z_l(\mathbf{r}-\mathbf{r_k})\}$, we approximate this coronagraph effect by putting weights $(c_{k,l})_{k\in \lbrack 1,n_{seg} \rbrack}$ on the coefficients $(a_{k,l})_{k\in \lbrack 1,n_{seg} \rbrack}$. These coefficients need to be computed. Therefore, the expression of the analytical model becomes:
\begin{equation} \label{eq:3}
    I(\mathbf{u}) = \left \Vert \widehat{Z_l}(\mathbf{u}) \right \Vert ^2 (\sum_{k=1}^{n_{seg}} c_{k,l} ^2 a_{k,l} ^2 + \sum_{i=1}^{n_{seg}} \sum_{j=1,j \ne i}^{n_{seg}} c_{i,l} a_{i,l} c_{j,l} a_{j,l} \cos((\mathbf{r_{j}} - \mathbf{r_{i}} ). \mathbf{u})),
\end{equation}
In the non-redundant pair-configuration, we obtain:
\begin{equation} \label{eq:27}
    I(\mathbf{u}) = \left \Vert \widehat{Z_l}(\mathbf{u}) \right \Vert ^2 (\sum_{k=1}^{n_{seg}} c_{k,l} ^2 a_{k,l} ^2 + 2 \sum_{q=1}^{n_{NRP}} A'_q \cos(\mathbf{b_q} . \mathbf{u})),
\end{equation}
where, for $q\in \lbrack 1,n_{NRP} \rbrack$, $A'_q = \sum_{(i,j)} c_{i,l} a_{i,l} c_{j,l} a_{j,l}$ and the couples $(i,j)$ are all the couples that verify the relation $\mathbf{r_{j}}-\mathbf{r_{i}}=\pm \mathbf{b_q}$.

- In a general case, PASTIS aims at computing the contrast. Hereafter, $\langle f \rangle_{DH}$ corresponds to the mean value in the dark hole of the image $f$. If we consider the general expression of the intensity of Eq.~\ref{eq:100}, we notice that 1) $\langle \left \Vert \mathcal{C}\{P\} \right \Vert ^2 \rangle_{DH}$ provides the deep contrast $C_0$ of the coronagraph and 2) $\langle 2 \Re{\{\mathcal{C}\{P\}\mathcal{C}\{\phi\}^*\}} \rangle_{DH} = 0$ in a symmetrical dark hole (explanations in the appendix~\ref{a_linear}). Therefore the contrast in the dark hole $C$ only takes into account the deep contrast $C_0$ and the average of the quadratic term $\langle \left \Vert \mathcal{C}\{\phi\} \right \Vert ^2 \rangle_{DH}$. More explanations in this direction are provided in the next section.



In practice, we calibrate the coefficients for each Zernike by running the end-to-end simulation with the chosen coronagraph and the analytical model without calibration once for each segment:
\begin{equation}
    \forall k\in \lbrack 1,n_{seg} \rbrack, c_{k,l} = \sqrt{\frac{\langle I_C(k)\rangle_{DH}}{\langle I(k)\rangle_{DH}}},
\end{equation}
where $I_C(k)$ is the intensity image issued from the end-to-end simulation after applying a $1$nm aberration of the $l$-th Zernike polynomial on the $k$-th segment only, and $I(k)$ is the intensity image produced by the analytical model after applying exactly the same phase to the segments.

This calibration has to be performed only once and integrates the coronagraph effect in the focal plane into our model.

\subsection{Matrix-based model}
\label{sec:M-AM}

As explained in the previous section, the deep contrast $C_0 = \langle \left \Vert \mathcal{C}\{P\} \right \Vert ^2 \rangle_{DH}$ can be taken into account. However we prove in the appendix~\ref{a_linear} that the average of the linear term, $\langle 2 \Re{\{\mathcal{C}\{P\}\mathcal{C}\{\phi\}^*\}} \rangle_{DH}$ is null in a symmetrical dark hole. Therefore the mean contrast in the dark hole is:
\begin{equation}
    C = C_0+\sum_{i=1}^{n_{seg}} \sum_{j=1}^{n_{seg}} c_{i,l} a_{i,l} c_{j,l} a_{j,l} \langle \left \Vert \widehat{Z_l}(\mathbf{u}) \right \Vert ^2 \cos((\mathbf{r_{j}} - \mathbf{r_{i}} ). \mathbf{u}) \rangle_{DH}
\end{equation}
This new expression can be expressed as a matrix product:
\begin{equation} \label{eq:13}
    C = C_0+a_l M_l a_l^t,
\end{equation}
where $a_l = (a_{1,l}, ..., a_{n_{seg},l})$ is the vector of the Zernike coefficients, $a_l^t$ is the transpose vector of $a_l$, and $\forall (i,j)\in \lbrack 1,n_{seg} \rbrack ^2, M_l(i,j) = c_{i,l} c_{j,l} m_{i,j,l}$, where $m_{i,j,l} = \langle \left \Vert \widehat{Z_l}(\mathbf{u}) \right \Vert ^2 \cos((\mathbf{r_{j}} - \mathbf{r_{i}} ). \mathbf{u}) \rangle_{DH}$.

This matrix-based version of the analytical model further reduces the computation time of the integrated contrast over a dark hole.

In the next section, the likelihood between the outputs issued from the PASTIS model and from the end-to-end simulation is quantified.

\section{Results}
\label{sec:Results}

In this section, we seek to validate the analytical model of the previous section, using a comparison with an end-to-end simulation. The chosen pupil and the coronagraph are described in section~\ref{sec:Choice of pupil and end-to-end simulation} and the results of comparison in sections~\ref{sec:M-AM results} and~\ref{sec:Application}.

\subsection{End-to-end simulations}
\label{sec:Choice of pupil and end-to-end simulation}

For this test, we choose the example pupil illustrated on Fig.~\ref{fig:Segmented_Pupils} (b), which is under study for future space telescopes in the Segmented Coronagraph Design and Analysis (SCDA) program\cite{Zimmerman2016}. It is formed of 36 identical hexagonal segments and a hexagonal central obstruction \cite{Postman2012,Feinberg2014}.

\begin{figure}
   \begin{center}
   \begin{tabular}{c}
      \includegraphics[height=3.5cm]{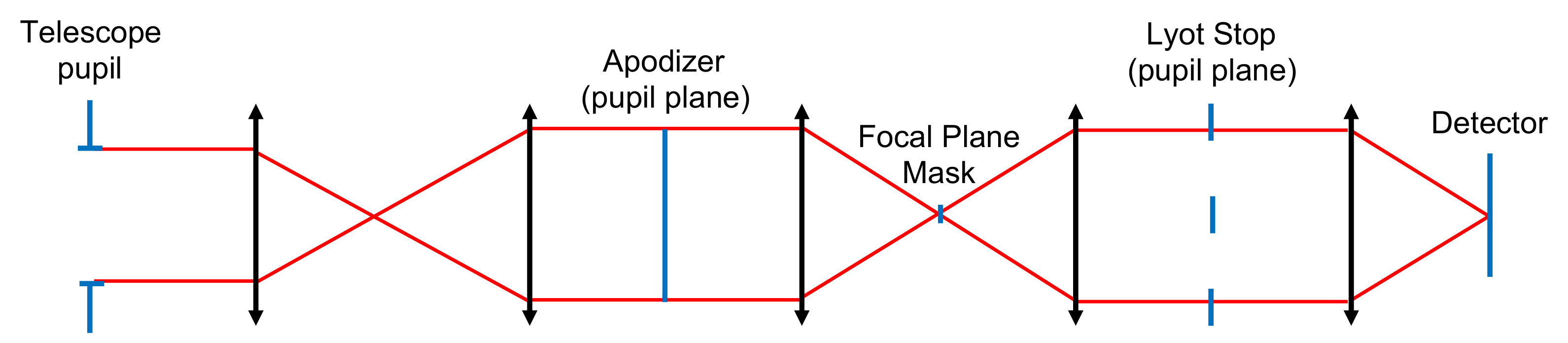} \\
   \includegraphics[height=5.3cm]{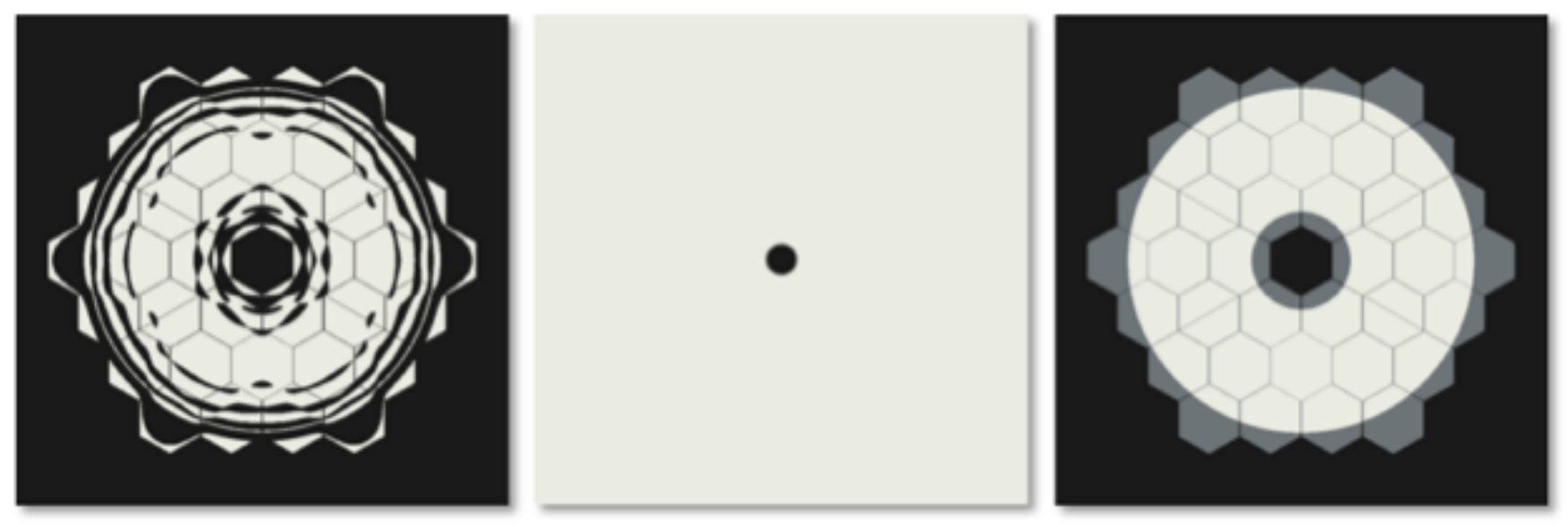}
   \end{tabular}
   \end{center}
   \caption[APLC_ATLAST] 
   { \label{fig:APLC_ATLAST} 
Top: Optical configuration of the APLC as used in the end-to-end simulation. Bottom: Optical masks used in the end-to-end simulation. The apodizer (left) is located in the first pupil plane, the focal plane mask (center) on the following focal plane, its radius here being $4.5 \lambda /D$, and the Lyot Stop (circular aperture on the right, here superposed with the entrance pupil) on the last pupil plane.}
   \end{figure}

\begin{figure}
   \begin{center}
   \begin{tabular}{c}
   \includegraphics[height=5cm]{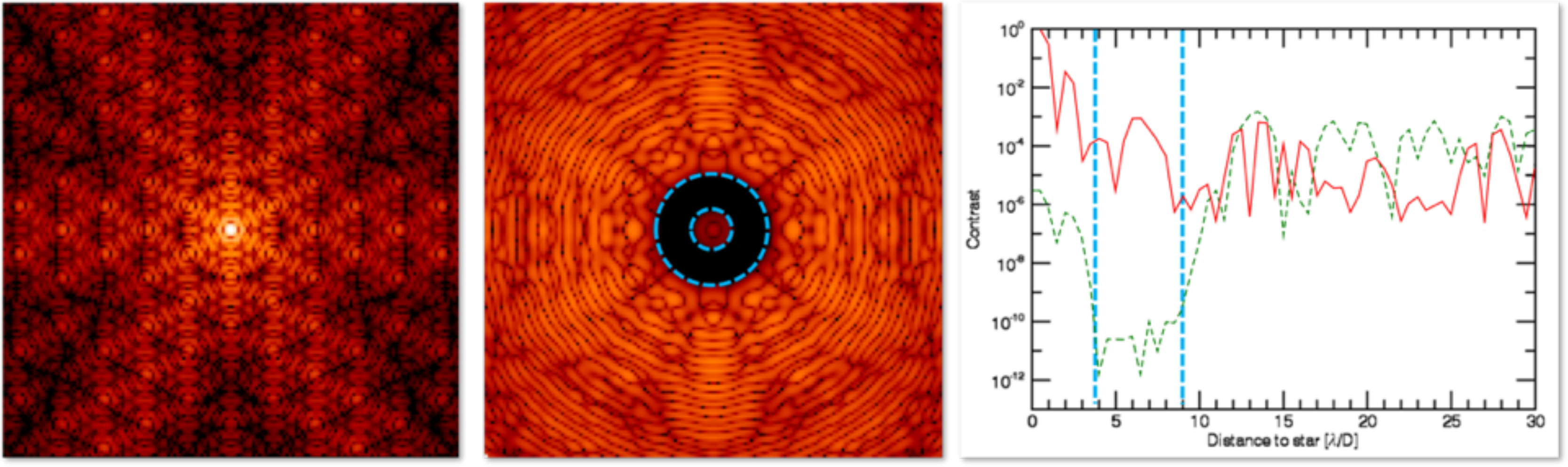}
   \end{tabular}
   \end{center}
   \caption[PSF_ATLAST] 
   { \label{fig:PSF_ATLAST} 
Left: Point Spread Function (PSF) in presence of the SCDA pupil of Fig.~\ref{fig:Segmented_Pupils}, with an end-to-end numerical simulation without coronagraph and without aberration. Center: PSF in presence of the same SCDA pupil combined with the APLC, with no aberration. Right: Cut along the horizontal radius of the two previous PSFs (red: without APLC, green: with APLC). We can observe that the APLC brings a huge correction in the dark hole, delimited here by the blue dashed lines at $4\lambda/D$ and $9\lambda/D$. The average contrast in this region is a few $10^{-11}$.}
   \end{figure}

The end-to-end simulation explicitly computes the propagation of the electric field from plane to plane with a Fourier formalism, using the functions of Paul et al\cite{Paul2014}. The coronagraph used in the end-to-end simulation is an Apodized Pupil Lyot Coronagraph (APLC) \cite{Soummer2003,NDiaye2015a,NDiaye2016ApJ}, specially designed for this pupil. The APLC, whose components are shown on Fig.~\ref{fig:APLC_ATLAST}, is designed to enable an extremely high contrast of a few $10^{-11}$ in the dark hole, which corresponds to a circular area between $4\lambda/D$ and $9\lambda/D$ (see Fig.~\ref{fig:PSF_ATLAST}). The analytical model is valid only in the high-contrast area, since $\widehat{P}(\mathbf{u})$ is far from 0 out of this area and can then still not be neglected anymore. For this reason, we look at the intensity images and performance in this area only and compare them to the outputs of the model.

\subsection{Results and comparison of the matrix-based analytical model}
\label{sec:M-AM results}

In this section, we compare the mean contrasts in the dark hole computed from the images of the end-to-end simulation and from the matrix-based analytical model.

\begin{figure}
   \begin{center}
   \begin{tabular}{c}
   \includegraphics[height=9cm]{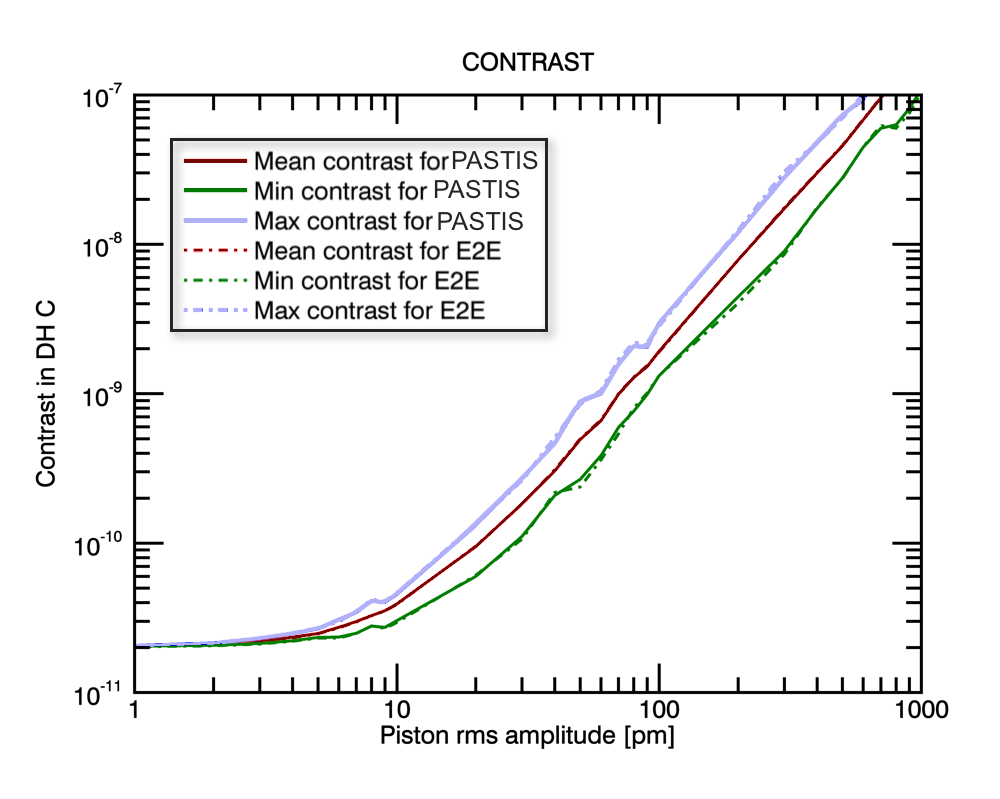}
   \end{tabular}
   \end{center}
   \caption[EB_contrast_AMvsE2E] 
   { \label{fig:EB_contrast_AMvsE2E} 
Contrast as a function of the rms piston error phase on the pupil, computed from both the end-to-end simulation (E2E) and PASTIS. This plot illustrates two regimes: below 10pm rms the contrast is limited by the coronagraph only and over a few 10pm rms the contrast is limited by the aberrations and the quadratic term is majority in the analytical model.}
   \end{figure}

Fig.~\ref{fig:EB_contrast_AMvsE2E} compares the output contrasts computed by the end-to-end simulation and from the matrix-based analytical model for piston aberrations from 1pm to 10nm rms on the segments. For each rms value, we select 250 random phases and compute the mean, minimum, and maximum contrasts over the 250 output contrasts. As a comparison, highly similar curves obtained thanks to end-to-end simulations only can be found in the studies of Stahl et al.\cite{Stahl2016b}, later completed in Stahl et al.\cite{Stahl2016c}.

The curves of Fig.~\ref{fig:EB_contrast_AMvsE2E} issued from PASTIS are highly similar to the ones issued for the end-to-end simulation. To quantify the error between the end-to-end simulation and PASTIS, we obtain an error in the contrast of around 3$\%$ rms.

However, PASTIS shows its best interest in such heavy computations, since its contrasts have been $10^7$ times faster to compute than the contrasts issued from the end-to-end model. Indeed PASTIS is a double multiplication between an $n_{seg}$-long vector, an $n_{seg}\times n_{seg}$-matrix, and another $n_{seg}$-long vector. Therefore its computation complexity is of the order of $n_{seg}^2$. On the other hand, one single end-to-end simulation requires three Fourier Transforms, which dominate the computation time by being of the order of $n\log(n)$, where $n$ is the size of the considered image. To plot figures with many iterations such as Fig.~\ref{fig:EB_contrast_AMvsE2E}, which has 203250 iterations, computation complexity has to be considered and PASTIS provides a significant gain of time and memory.

\subsection{Images generated with PASTIS}
\label{sec:Application}

PASTIS has been validated in the previous section, providing a fast estimation of the contrast with a $3\%$ error. In this section, we now look at the images obtained with the model before developing the matrix-based formula (equation~\ref{eq:27}) to compare the morphology of the speckles in the images themselves.

To do so, we focus on the cases where only local low-order aberrations are applied on the segments: piston, tip, tilt, focus, and the two astigmatisms. Furthermore, for each Zernike polynomial, two different configurations are compared: a case where two segments only have local aberrations and a case where random low-order Zernike coefficients are applied on all the segments.

\begin{figure}
   \begin{center}
   \begin{tabular}{cc}
   \includegraphics[height=10cm]{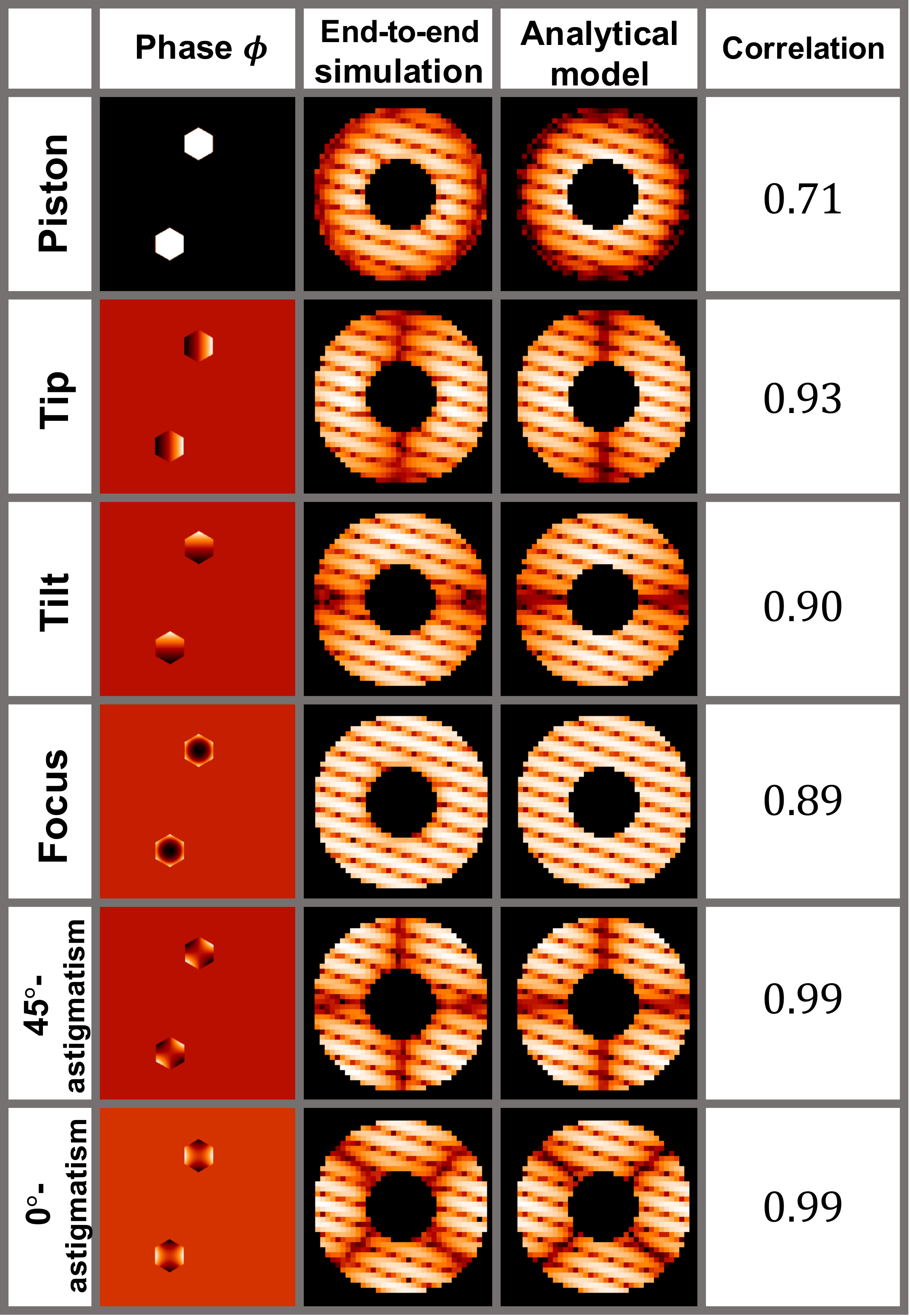} &
   \includegraphics[height=10cm]{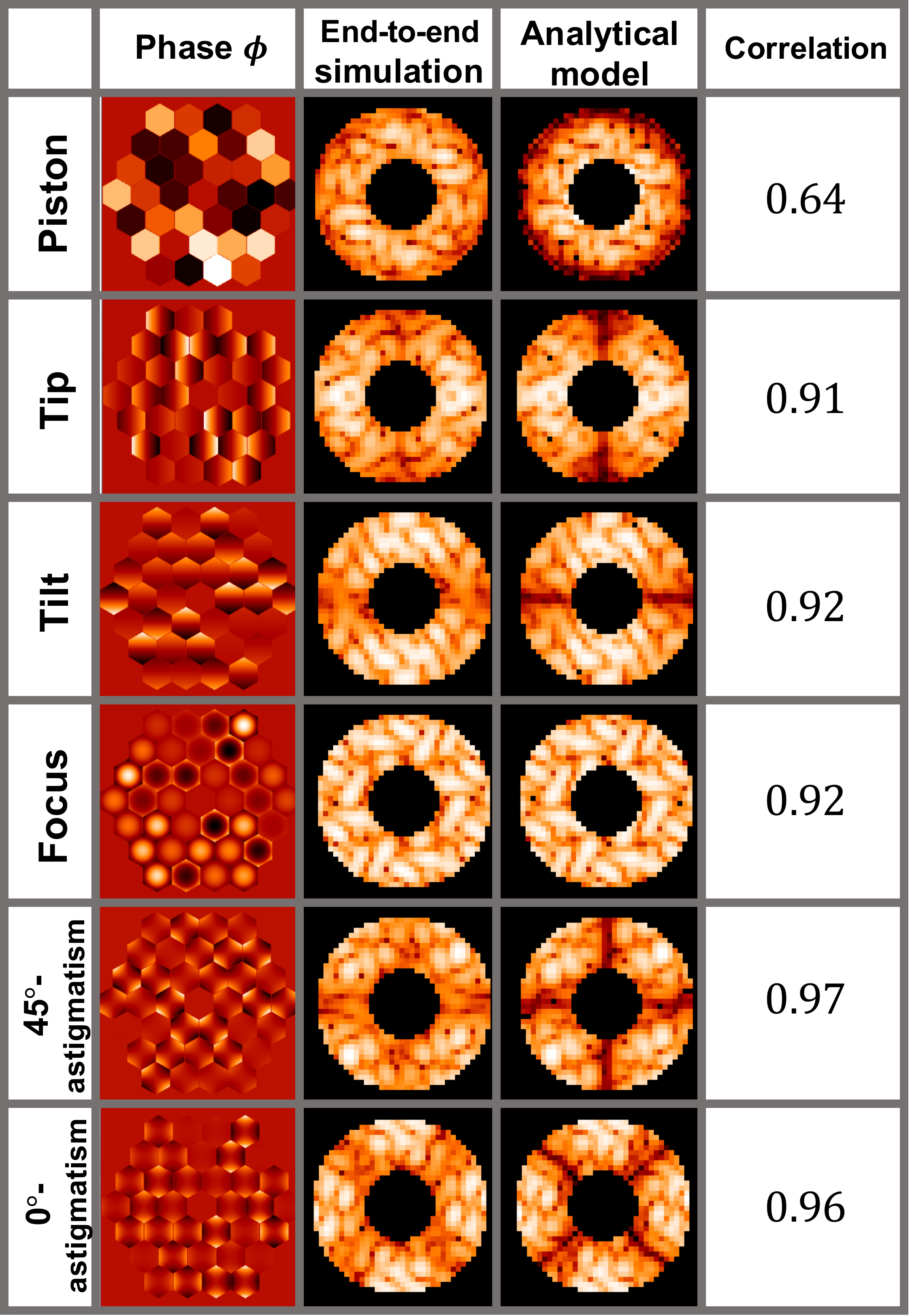} \\
   (a) & (b)
   \end{tabular}
   \end{center}
   \caption[PSF] 
   { \label{fig:PSF} 
Comparison of the PSFs (log scale) when (a) identical 1nm aberrations are applied on a pair of segments and when (b) random aberration coefficients are applied on the segments for the first six Zernike polynomials (piston, tip, tilt, focus, and the astigmatisms). A binary mask is applied on the PSFs to display the dark hole only. The first column indicates the phase applied on the primary mirror. The PSFs of the second column are generated by a simulation of an end-to-end propagation through an APLC. The PSFs of the third column are generated by the analytical model. The last column gives the linear Pearson correlation coefficients between the end-to-end simulation and analytical model images. All the horizontal lines are at the same scale.}
   \end{figure}

The PSFs resulting from the end-to-end simulation and the ones resulting from the analytical model are indicated in Fig.~\ref{fig:PSF}. All panels are at the same scale, which illustrates once again that the analytical model provides a good prediction of the overall contrast level. We then use the correlations to quantify the agreement on the morphology of the speckles' intensity between the analytical model and the end-to-end simulations. The correlation here corresponds to the linear Pearson correlation coefficient, which, for two images $I$ and $J$ of $N$ elements, is computed as:
\begin{equation}
    Cor(I,J) = \frac{\sum_{k=1}^{N}(I(k)-\langle I \rangle)(J(k)-\langle J \rangle)}{\sqrt{\sum_{k=1}^{N}(I(k)-\langle I \rangle)^2 \sum_{k=1}^{N}(J(k)-\langle J \rangle)^2}}
\end{equation}
In the piston case, the images typically have correlations of around $70\%$ and in the other cases, typically closer than $89\%$.

As a reminder, the envelopes of these different Zernike polynomials are shown in Fig.~\ref{fig:enveloppes}. In the piston case, it generates a circle at the limit of the dark hole. In the tip and tilt cases, they create a dark vertical or horizontal line crossing the center of the PSF. In the focus case, the envelope has a ring shape. Finally, in both astigmatisms, the envelopes create a cross in the dark hole. The effects of these different envelopes clearly appear in the images issued both from the end-to-end simulation and from the analytical model. On Fig.~\ref{fig:PSF}, the envelopes also seem the main source of the difference between the PSFs issued from the analytical model and from the end-to-end simulation. This effect might be an artefact due to the apodization of the pupil: the envelopes were computed with a Zernike polynomial defined on a regular hexagonal segment, while the support shape depends on the apodization.

Another source of error is the scalar and the linear terms of Eq.~\ref{eq:100}, that are not taken into account in the model and therefore not displayed here.

This comparison between end-to-end simulation images and the outputs from the analytical model indicates that PASTIS allows to generate high-contrast images with a high-fidelity average contrast. Even if the morphologies of the images do not seem extremely accurate, mainly in the piston case (correlation around $70\%$, while over $90\%$ for the other Zernike polynomials), the contrast values are close enough for error budgeting.


In practice, we keep using the matrix-based analytical model, since it enables a fast estimate of the contrast in the dark hole, which is our chosen criterion for error budgeting, without the need of an actual image.



\section{Sensitivity analysis}
\label{sec:Sensitivity Analysis}

A traditional error budget aims at quantifying the deterioration of the contrast with the rms error phase applied on the segments. For example, in the piston case, we can easily deduce from Fig.~\ref{fig:EB_contrast_AMvsE2E} the constraints in piston cophasing in term of rms error. For instance, to obtain a contrast of $10^{-10}$, the piston phasing needs to be accurate to better than 20pm rms.

PASTIS is a faster method to compute the deterioration of contrast due to segment-level aberrations. In particular, it can speed up large forward Monte-Carlo to translate multiple realizations of deployment, phasing, or wavefront control. Similarly, it makes simulations of performance for long-time series of high-frequency vibrations possible. The most direct way to use PASTIS to find requirements for a given contrast is to apply it multiple times with multiple realizations of multiple errors, as shown in Fig.\ref{fig:EB_contrast_AMvsE2E}. However, we have then no clue of the repartition of these requirements on the segments, while it is known that some segments have a bigger impact on the contrast than others: for example, the calibration coefficients $(c_{k,l})_{k\in \lbrack 1,n_{seg} \rbrack}$ depend on the segment. PASTIS can actually identify these segments or group of segments. Using these modes, doing an error budget or designing wavefront sensing and control systems is a lot easier. 

This is why we develop a sensitivity analysis method based on a modal projection of the phase, which takes into account this segment-dependant contribution to the contrast.

\subsection{Analytical modal inversion}
\label{sec:Modal inversion}

In equation~\ref{eq:13} of section~\ref{sec:M-AM}, the contrast for the $l$-th Zernike depends on the matrix $M_l$. $M_l$ being symmetric and positive-definite, there exist $U_l$, a matrix of eigen vectors, and $D_l$, a diagonal matrix of eigen values, that satisfy:
\begin{equation}
    D_l = U_l M_l U_l^t
\end{equation}

We call $(\lambda_{p,l})_{p\in \lbrack 0,n_{seg}-1 \rbrack}$ the eigen values of $M_l$ and $(U_{p,l})_{p\in \lbrack 0,n_{seg}-1 \rbrack}$ their associated eigen vectors. Therefore we have:
\begin{equation}
    \forall p\in \lbrack 0,n_{seg}-1 \rbrack, \lambda_{p,l} = U_{p,l} M_l U_{p,l}^t
\end{equation}

Since $(U_{p,l})_{p\in \lbrack 0,n_{seg}-1 \rbrack}$ forms a modal basis, the error phase segment coefficients $A$, which provides a final contrast $C$, can be expressed in an unique way as a function of these eigen modes:
\begin{equation}
    A = \sum_{p=0}^{n_{seg}-1} \sigma_p U_{p,l},
\end{equation}
where $\sigma_p$ is the contribution of the mode $p$.

The projection of the error phase $\Phi$ on each eigen mode $U_{p,l}$ contributes as much as $C_p$ to the final contrast. 

$(U_{p,l})_{p\in \lbrack 0,n_{seg}-1 \rbrack}$ forming a basis of orthonormal vectors, the final contrast is indeed the sum of the contrasts generated by the different projections on the modes:
\begin{equation} \label{eq:6}
    C = \sum_{p=0}^{n_{seg}-1} C_p
\end{equation}
Therefore, we have:
\begin{equation}
    \forall p\in \lbrack 0,n_{seg}-1 \rbrack, C_p = (\sigma_p U_{p,l}) M_l (\sigma_p U_{p,l})^t
\end{equation}
\begin{equation} \label{eq:5}
    \hspace{4em} = \sigma_p^2 \lambda_{p,l}
\end{equation}

As a conclusion, to get a contribution to contrast smaller than $C_p$ on the $p$-th mode, the projection of the phase on this mode has to be smaller than
\begin{equation} \label{eq:4}
    \sigma_p = \sqrt{\frac{C_p}{\lambda_{p,l}}}.
\end{equation}

\subsection{Strategy}
\label{sec:Strategy}

Hereafter, we define the contrast as a sum of a static contribution and a dynamic contribution:
\begin{equation}
    C = C^s \pm \Delta C.
\end{equation}

We assume that the coronagraphic contrast is of the order of $10^{-10}$ in the dark hole, but we require system stability sufficient to support improving the contrast, to $10^{-10}$. Therefore, we make the assumptions that
\begin{equation}
    \begin{cases}
    C^s = 10^{-10} \\
    \Delta C = 10^{-10}
    \end{cases}
\end{equation}

Furthermore, the assumption is made that all the modes (except for the one with an extremely low eigen value) contribute equally and independently to the contrast and its stability. Therefore, we formulate the hypotheses:
\begin{equation}
    \begin{cases}
    C_p^s = \frac{C_p^s}{35} \\
    \Delta C_p = \frac{\Delta C}{\sqrt{35}}
    \end{cases}
\end{equation}

The formula~\ref{eq:4} provides both the mode contributions $(\sigma_p)_{p\in \lbrack 0,n_{seg}-1 \rbrack}$ that would generate such contrasts on each mode, and the mode contributions in terms of stability $(\Delta \sigma_p)_{p\in \lbrack 0,n_{seg}-1 \rbrack}$.

The latter sections impose the system stability requirements for piston and $45^\circ$-astigmatism.

\subsection{Illustration in the case of local pistons in the pupil}
\label{sec:Application modal}

\subsubsection{Eigen modes}

\begin{figure}
   \begin{center}
   \begin{tabular}{c}
   \includegraphics[height=9cm]{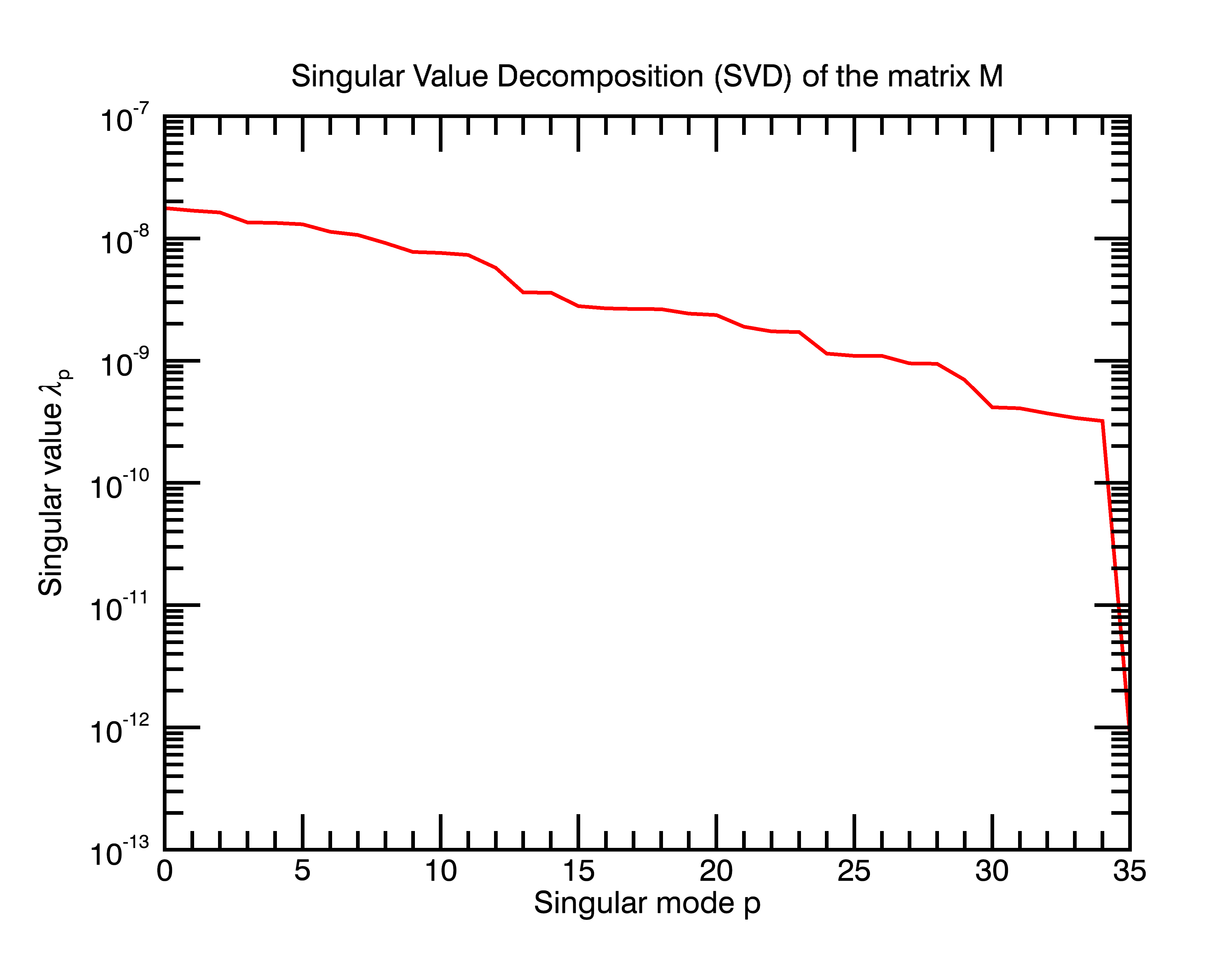}
   \end{tabular}
   \end{center}
   \caption[SingularValues] 
   { \label{fig:SingularValues} 
Eigen values of the matrix $M_1$ in the segment-level piston case. The last eigen value, extremely lower than the others, corresponds to a eigen mode of a global piston on the primary mirror.}
   \end{figure}

\begin{figure}
   \begin{center}
   \begin{tabular}{c}
   \includegraphics[height=7cm]{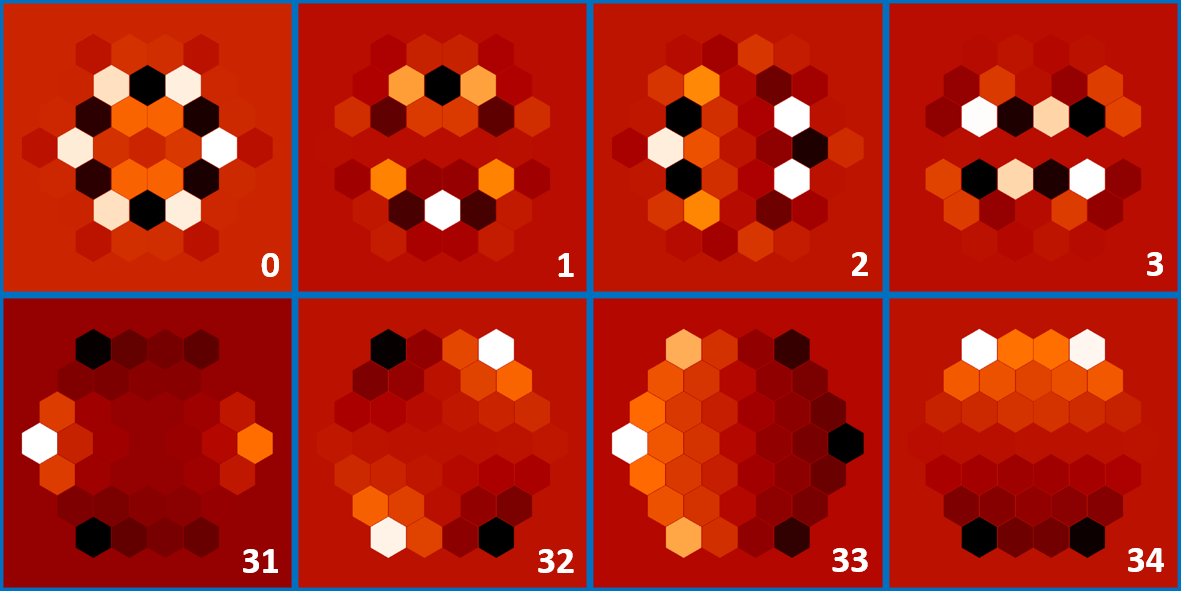}
   \end{tabular}
   \end{center}
   \caption[SomeModes] 
   { \label{fig:SomeModes} 
Eigen modes in the local only piston case. The top line corresponds to the four modes with the highest eigen values, the bottom line to four of the modes with the lowest eigen values. In this second line, we can recognize discrete versions of some common low-order Zernike polynomials: the two astigmatisms and the tip and tilt. Furthermore, the last modes focus more on the corner segments, that are typically the segments that impact the contrast the least, since they are the most obscured by both the apodizer and the Lyot stop. Conversely, on the top line, we can also see that the segments with the most extreme piston coefficients correspond to the segments hidden by neither the apodizer nor the Lyot stop, and so are the segments that influence the contrast the most. This explains why they have the highest eigen values.}
   \end{figure}

In the piston case, $C = a_1 M_1 a_1^t$. A Singular Value Decomposition (SVD) is applied to the matrix $M_1$, the eigen values are indicated in Fig.~\ref{fig:SingularValues}. Fig.~\ref{fig:SomeModes} indicates a few eigen modes computed from this SVD: on top, the four modes attached to the highest eigen values of $M_1$, and on the bottom, the four modes attached to low eigen values. The very last mode, which has an extremely low eigen value, is not indicated here and corresponds to a global piston on the primary mirror.

From this SVD, we can see that the last eigen modes correspond to discretized global low order Zernike polynomials, the two astigmatisms (modes 31 and 32) and the tip and tilt (modes 33 and 34). We find that the lowest sensitivity modes of our simplified model are the discretized low-order Zernike modes tip and tilt, which the APLC has been designed to be robust to. This consolidates the realism of our analytical model PASTIS. Furthermore, the very last mode, not indicated in Fig.~\ref{fig:SomeModes}, is a global piston on the entire pupil, which has here an extremely low eigen value and is known not to affect the contrast in a realistic coronagraph. The highest eigen values correspond to the modes that affect the contrast the most, a combination of segments of the second ring which are known to be the least apodized and the least hidden by the Lyot Stop. In particular, we can conclude that these modes are specific to the chosen coronagraph.

\subsubsection{Constraints on phasing}

The hypotheses and strategy to compute the static constraints on the modes are indicated in section~\ref{sec:Strategy}.

\begin{figure}
   \begin{center}
   \begin{tabular}{cc}
   \includegraphics[height=6cm]{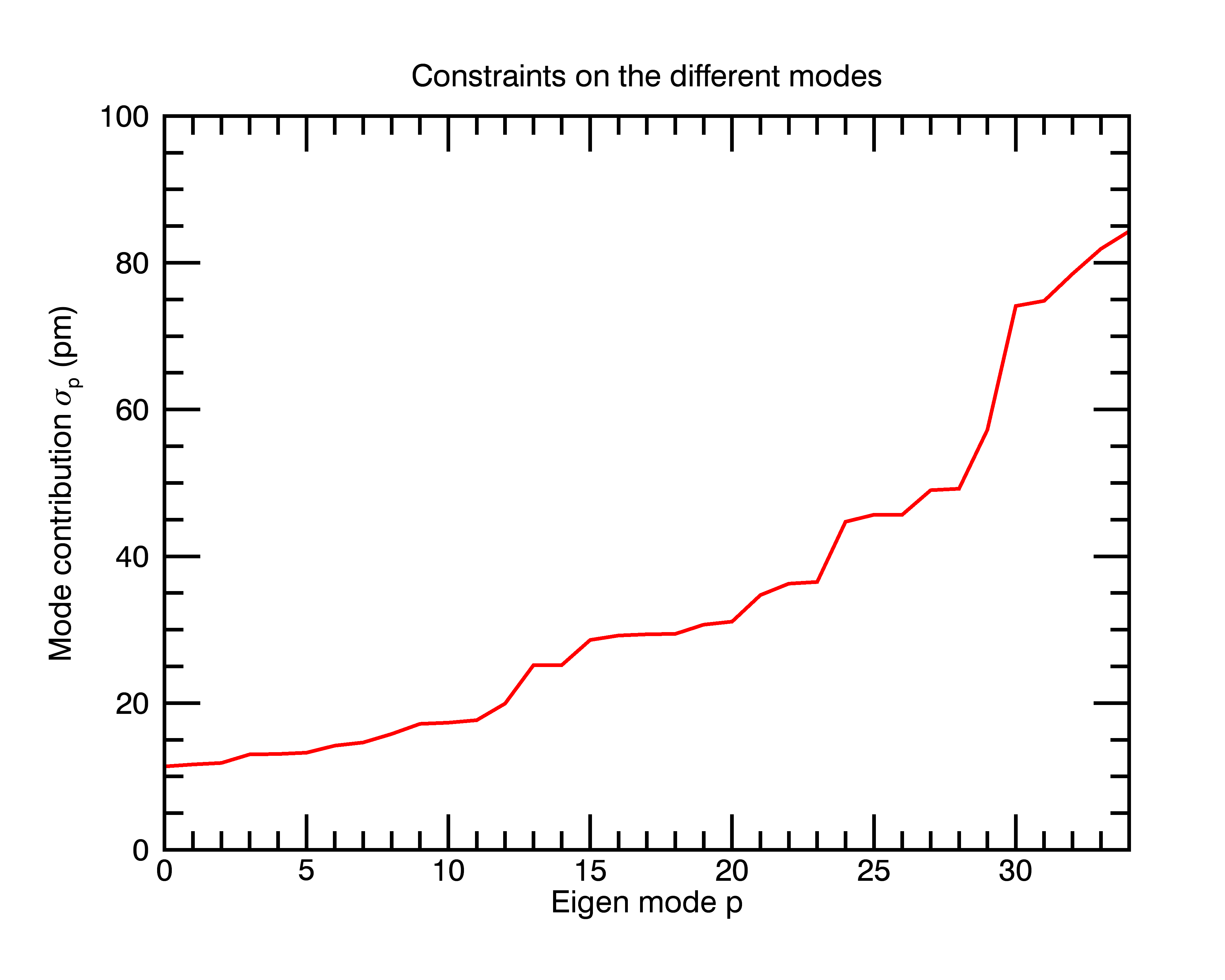} & \includegraphics[height=6cm]{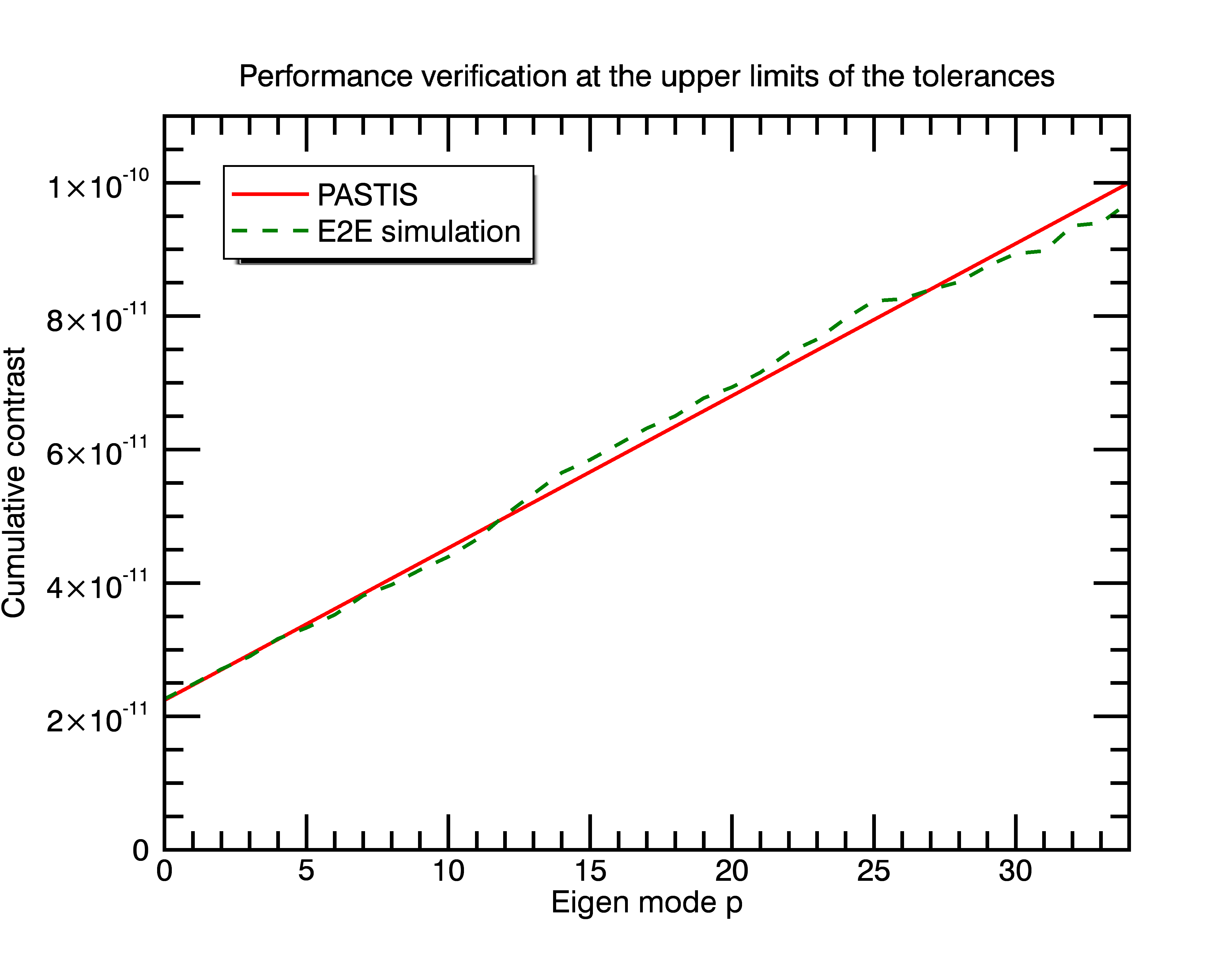}
   \end{tabular}
   \end{center}
   \caption[Constraints_1e-6_SVDInversion_Piston] 
   { \label{fig:Constraints_1e-6_SVDInversion_Piston} 
Left: Contributions $(\sigma_p)_{p\in \lbrack 0,n_{seg}-1 \rbrack}$ on the different piston modes to reach a final target contrast of $10^{-10}$, in the case where only local pistons on segments deteriorate the contrast. Right: Cumulative contrasts on these piston modes at their upper constraints to reach a final target contrast of $10^{-10}$. In these two plots, only 35 modes are indicated, since the mode with a very low eigen value corresponds to a global piston on the pupil and is chosen to not contribute to the final contrast.}
   \end{figure}

After applying the equation~\ref{eq:4}, we obtain the results indicated in Fig.~\ref{fig:Constraints_1e-6_SVDInversion_Piston}: on the left, the mode contributions $(\sigma_p)_{p\in \lbrack 0,n_{seg}-1 \rbrack}$ that generate such contrasts are indicated and on the right, the cumulative contrasts generated by these constraints are shown. Each mode contributes equally to the contrast and the final contrast is $C = 10^{-10}$, which were our hypotheses. We also indicated here the cumulative contrasts when these constraints are injected as inputs of the end-to-end simulation, and despite an error on the final contrast of $3.75\%$, we can conclude that this method to compute the tolerances is relevant and useful. 

\begin{figure}
   \begin{center}
   \begin{tabular}{c}
   \includegraphics[height=9cm]{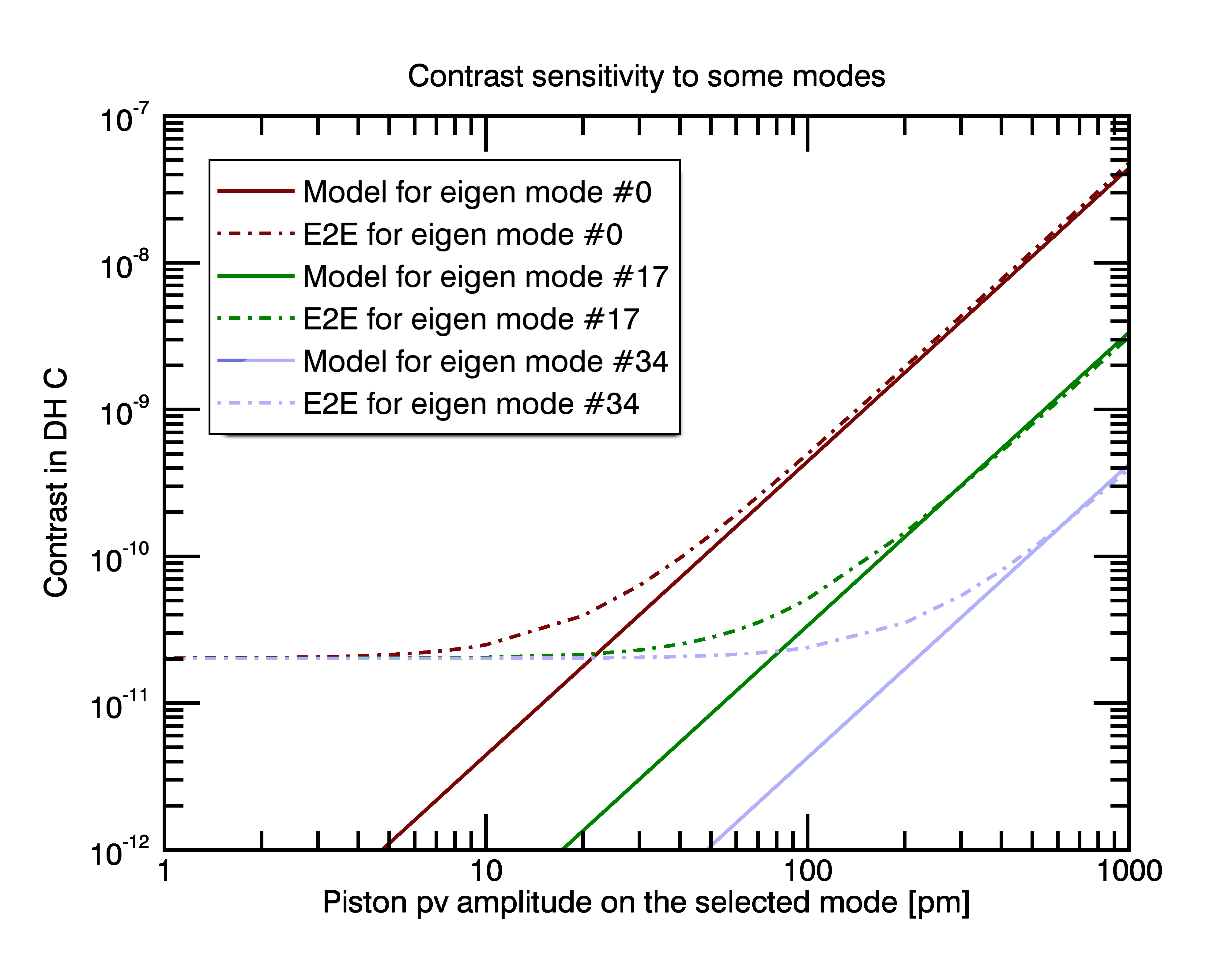}
   \end{tabular}
   \end{center}
   \caption[Piston_SVD_Inversion_SomeModes] 
   { \label{fig:Piston_SVD_Inversion_SomeModes} 
Contrast as a function of the amplitude on three different piston modes, for both the end-to-end simulation (E2E) and the matrix-based analytical model. The red curves correspond to the mode 0, which has the highest eigen value so the highest influence on the contrast. As a consequence, its constraints to reach a target contrast are tougher than the constraints obtained for any other mode. The green curves correspond to the intermediate piston mode 17, and the blue curves correspond to the mode with the lowest eigen value: as a consequence, its constraints are relaxed.}
   \end{figure}

In the general case, the mode contributions $(\sigma_p)_{p\in \lbrack 0,n_{seg}-1 \rbrack}$ both depend on the mode $p$ and on the contrast to reach in each mode $(C_p)_{p\in \lbrack 0,n_{seg}-1 \rbrack}$. Fig.~\ref{fig:Piston_SVD_Inversion_SomeModes} illustrates the contrast $C_p$ as a function of the contribution $\sigma_p$ for three different modes. The constraints clearly depend on the modes, the first modes requiring tighter constraints.

\subsubsection{Quasi-static stability}
\label{sec:Quasi-static stability}

In this section, the same inversion of the analytical model is applied to the contrast stability, which is ensured by the stability of the segment aberrations. The hypotheses and strategy were indicated in section~\ref{sec:Strategy}.

\begin{figure}
   \begin{center}
   \begin{tabular}{c}
   \includegraphics[height=9cm]{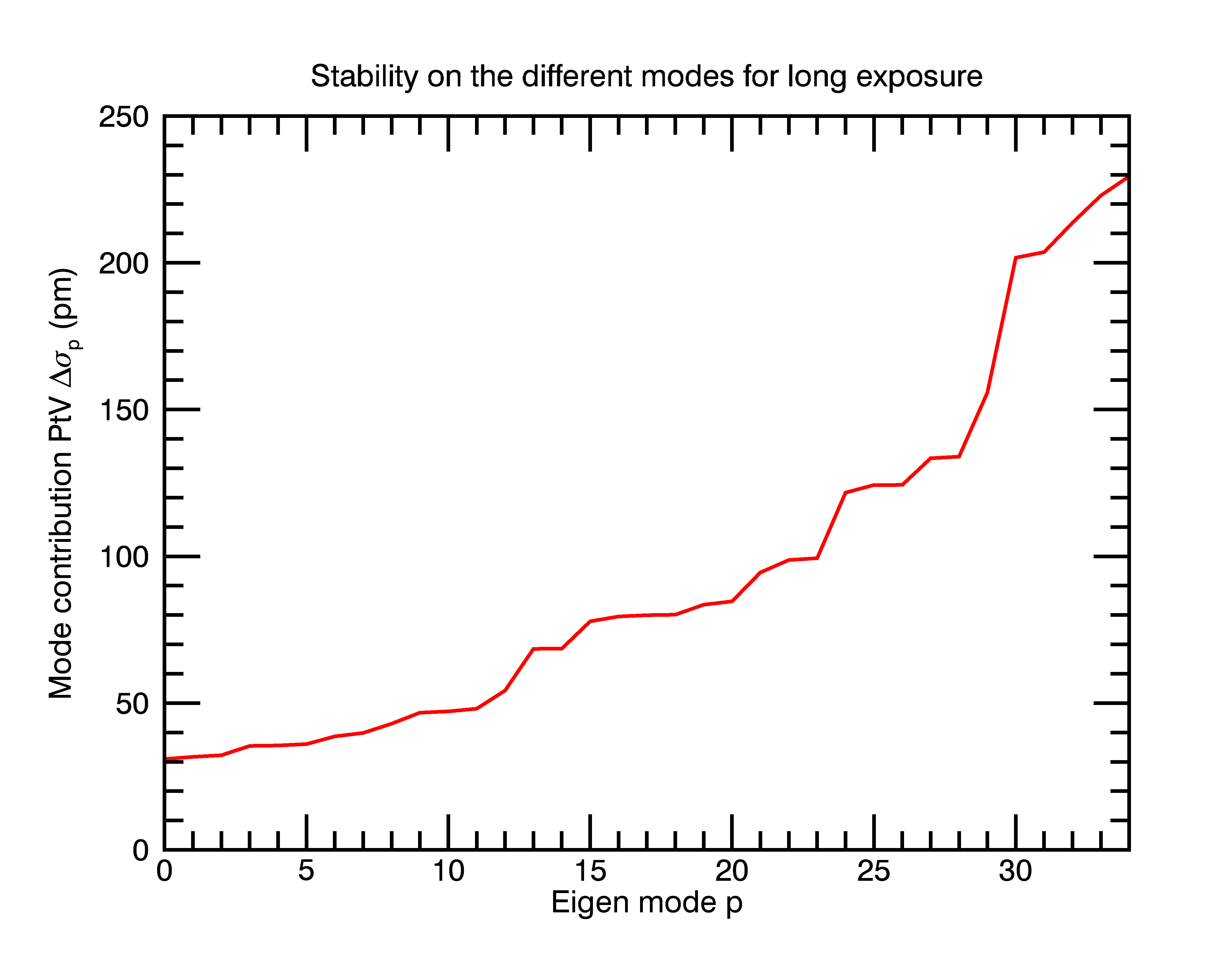}
   \end{tabular}
   \end{center}
   \caption[Stability_1e-6_SVDInversion_Piston] 
   { \label{fig:Stability_1e-6_SVDInversion_Piston} 
Stability coefficients on the different piston modes to reach a final target stability on contrast of $10^{-10}$ on a long exposure, in the case where only local pistons on segments deteriorate the contrast.}
   \end{figure}

We then obtain the constraints per mode in stability indicated on Fig.~\ref{fig:Stability_1e-6_SVDInversion_Piston}. We mainly notice that the most constraining modes require a stability of around 30pm.

In this example, for a contrast better than $10^{-10}$ stabilized at $10^{-10}$, the constraints for the modes 0, 17, and 34 are indicated in table~\ref{table:contraintes}.

\begin{table}
\centering
\begin{tabular}{|c|c|c|c|}
  \hline
  Mode number $p$ & 0 & 17 & 34 \\
  \hline
  Maximum mode contribution $\sigma_p$ (pm) & 11.36 & 29.37 & 84.30 \\
  \hline
  Maximum stability mode contribution $\Delta \sigma_p$ (pm) & 30.93 & 79.95 & 229.47 \\
  \hline
\end{tabular}
\caption{Absolute and stability constraints on three different modes for a contrast better than $10^{-10}$ stabilized at $10^{-10}$.}
\label{table:contraintes}
\end{table}

\subsection{Illustration in the case of local $45^\circ$-astigmatisms in the pupil}

Now, we consider constraints on local $45^\circ$-astigmatisms on the segments. Once again, the performance to achieve in the dark hole with the coronagraph only is $10^{-10} \pm 10^{-10}$.

\begin{figure}
   \begin{center}
   \begin{tabular}{c}
   \includegraphics[height=7cm]{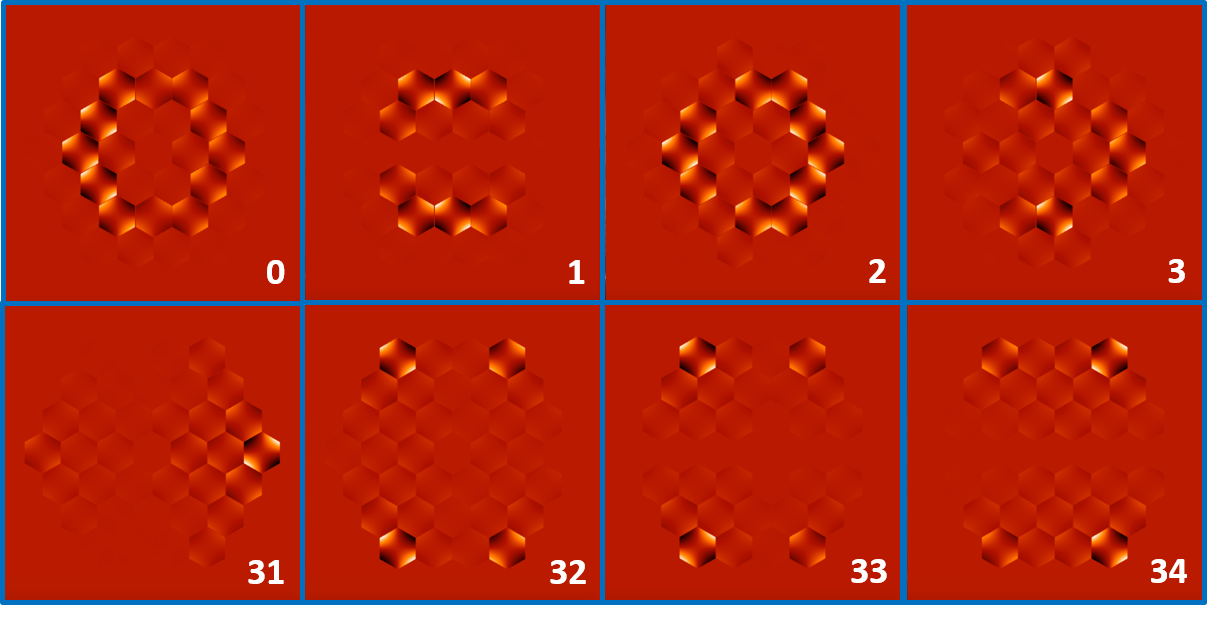}
   \end{tabular}
   \end{center}
   \caption[SomeModesAstig] 
   { \label{fig:SomeModesAstig} 
Eigen modes in the local only $45^\circ$-astigmatism case. The top line corresponds to the four modes with the highest eigen values, the bottom line to four of the modes with the lowest eigen values. On the top line, we can see that the segments with the most extreme $45^\circ$-astigmatism coefficients correspond to the segments hidden by neither the apodizer nor the Lyot stop, so the segments that influence the contrast the most, which explains why they have the highest eigen values. On the opposite, the modes of the second line focus on the corner segments only, which are the segments the most hidden by both the Lyot Stop and the apodizer.}
   \end{figure}

A few selected eigen modes are shown in Fig.~\ref{fig:SomeModesAstig}. Like in the piston case, the first eigen modes focus on the second ring, that is the only one not hidden by the Lyot Stop and the least apodized. On the opposite, the last modes correspond to the corner segments, which are both the most hidden by the Lyot Stop and the most apodized. We can conclude that the second ring, like in the piston case, require the tightest constraints in manufacturing and alignment and might require special consideration in backplane design. This conclusion would not have been formulated with a traditional error budget based on numerous end-to-end simulations. It is here shown that another advantage of the model is its ability to build a comprehensive analysis of the impact of each perturbation term.

\begin{figure}
   \begin{center}
   \begin{tabular}{cc}
   \includegraphics[height=6.5cm]{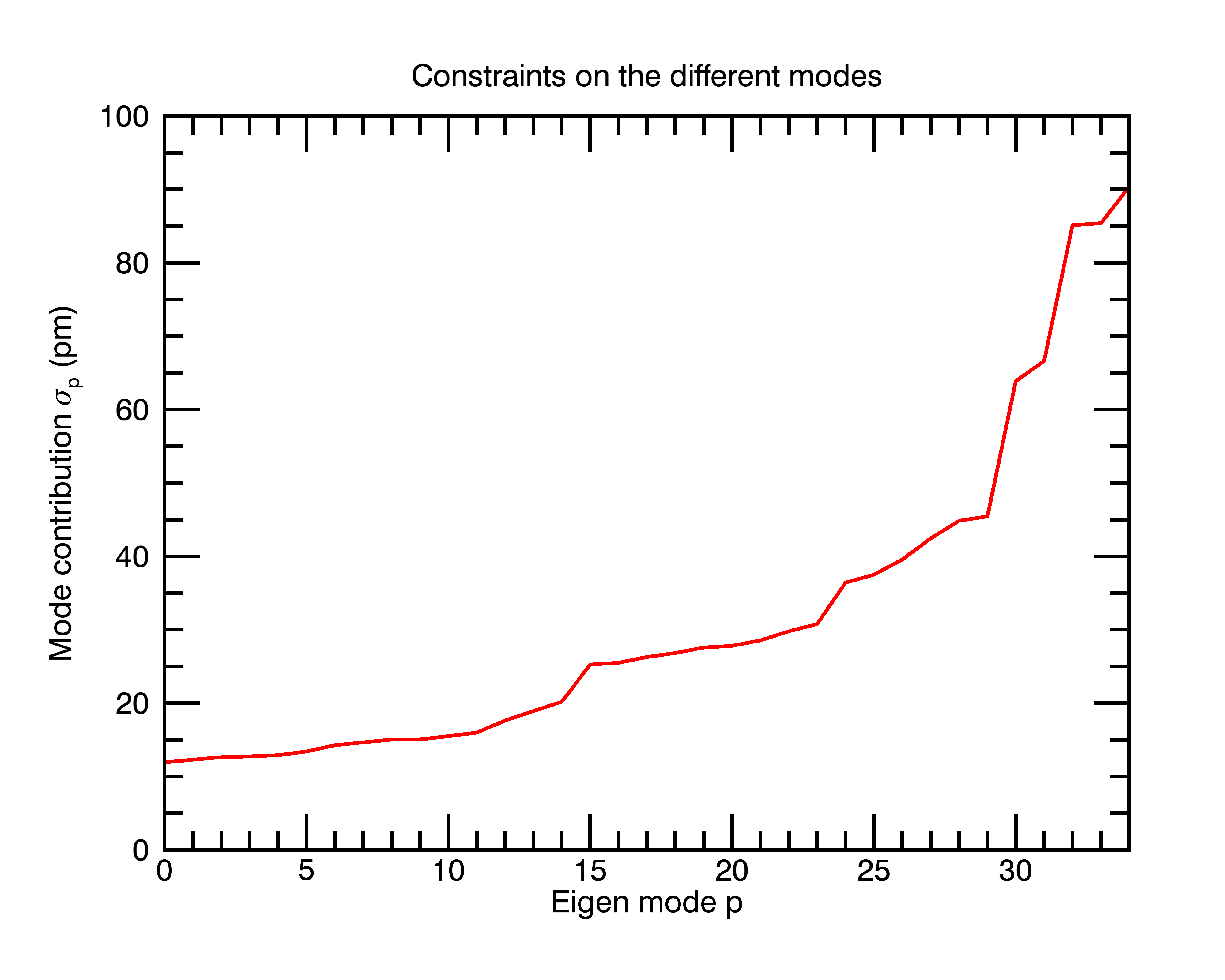}
    \includegraphics[height=6.5cm]{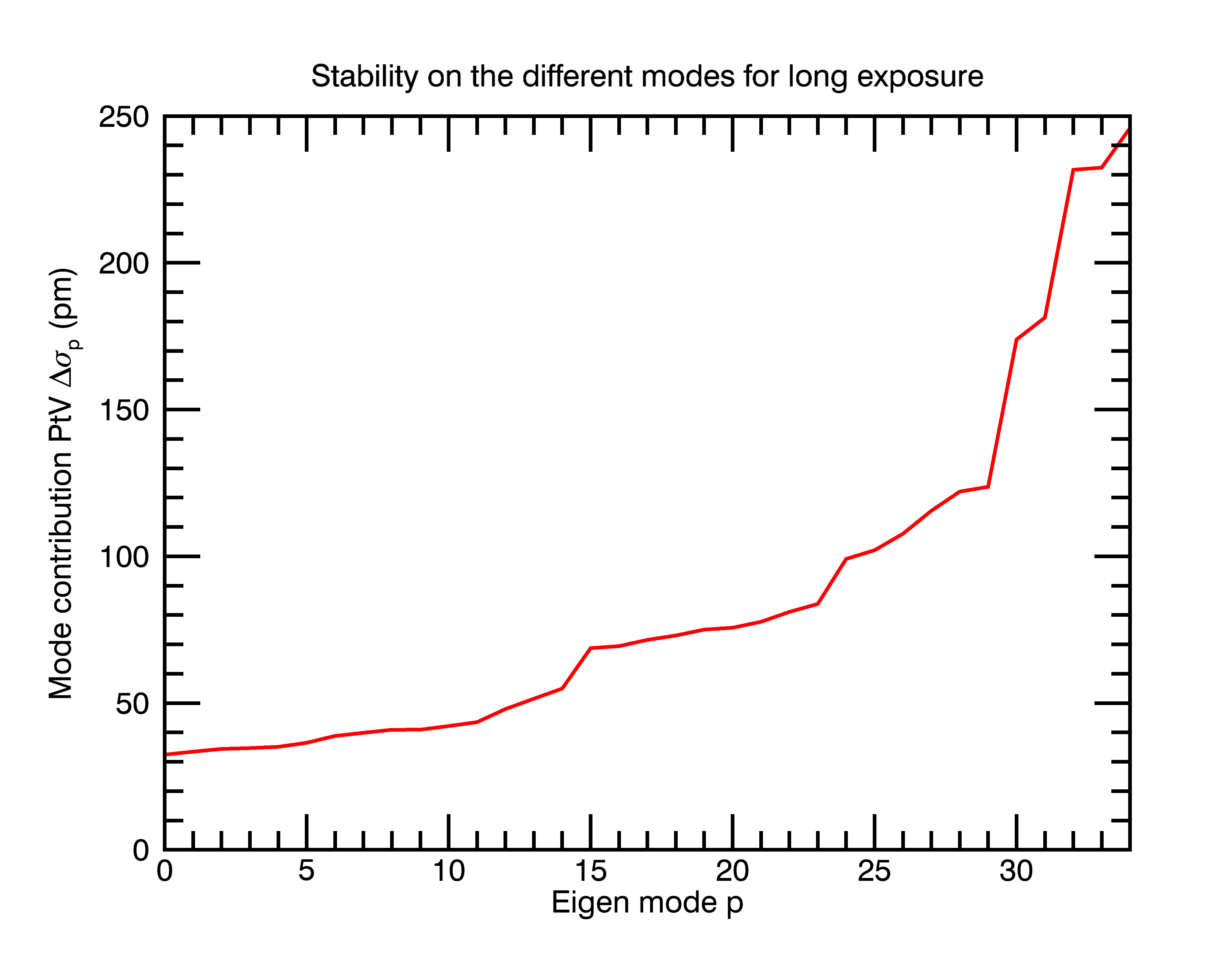}
   \end{tabular}
   \end{center}
   \caption[SVDInversion_Astig45] 
   { \label{fig:SVDInversion_Astig45} 
Left: Contributions $(\sigma_p)_{p\in \lbrack 0,n_{seg}-1 \rbrack}$ on the different $45^\circ$-astigmatism modes to reach a final target contrast of $10^{-10}$, in the case where only local $45^\circ$-astigmatisms on segments deteriorate the contrast. Right: Stability coefficients $(\Delta \sigma_p)_{p\in \lbrack 0,n_{seg}-1 \rbrack}$ on the different $45^\circ$-astigmatism modes to reach a final target stability of $10^{-10}$ on a long exposure.}
   \end{figure}

The numerical results in terms of sensibility analysis on the different modes are shown in Fig.~\ref{fig:SVDInversion_Astig45}. We can observe that the constraints are tighter or equal than the ones on piston on the first eigen modes, but looser on the last modes. 

These last two applications are just some examples of the use of the analytical model for sensitivity analysis. They also illustrate the gain in time and in understanding that such a formal inversion would provide compared to multiple end-to-end simulations in error budgeting. They can also provide a better understanding of the constraint geometry of the primary mirror, to design for example an adequate backplane, edge sensor placement, or laser truss geometry.

\section{Conclusions}
\label{sec:Conclusions}

In this paper we have introduced an analytical model, PASTIS, that is the basis of a new method for tolerancing of segmented telescopes with high-contrast instruments, both in term of segment alignment and stability. The formalism has been validated for one kind of local Zernike polynomial at a time on the segments, using comparisons between its outputs and the images and contrasts determined from the simulation of an end-to-end propagation through a coronagraphic system. The error between the contrasts computed from the analytical model and from the end-to-end simulation is around $3\%$ rms. The clear advantage of the analytical model is the reduction of computation time: to compute a contrast in the dark hole, an end-to-end simulation takes around $10^7$ longer than the matrix-based analytical model PASTIS. Therefore, PASTIS enables complete error budget with a significant gain of time, since a traditional error budget is based on tremendous contrast computations from phases selected thanks to a Monte Carlo distribution.

However the primary advantage of this model is that the entire method presented here is based on a simple formal inversion based on a modal analysis. This new process provides an easy and fast way to estimate the tolerancing for a given contrast, but also the constraints in term of stability. It also provides a more comprehensive analysis of the system, with the eigen modes that affect the contrast the most and required a thinest adjustment. A better understanding of the critical modes in the primary mirror can for example enable an optimization of the architecture of the backplane or of the positions of edge sensors.

The next step of this study is a generalization of this analytical model to a combination of Zernike polynomials, to understand the Zernike to Zernike dependency to contrast value. Furthermore, the static or quasi-static errors on the segments are not the only issue in high-contrast imaging. The telescope vibrations or the resonant modes of the segments are a main source of instability, and therefore are important factors in the limitation of the performance. As soon as the temporal aspect is considered, a full and time-consuming end-to-end simulation becomes inconvenient, and PASTIS shows its best advantage. Finally, we plan to compare its accuracy on other coronagraphs, for example using a Vortex coronagraph instead of an APLC.

This new formalism to describe segmented pupils and generate images and contrasts is adaptable to any segmented pupils, such as the Extremely Large Telescopes, the Thirty Meters Telescopes, the James Webb Space Telescope or the new HabEx and LUVOIR pupils. It can also be applied to non-hexagonal-segment pupils, such as the Giant Magellan Telescope. PASTIS enables a new, fast, and efficient method for static error budget and stability analysis for all segmented telescopes. 

\acknowledgments 
This work is supported in part by the National Aeronautics and Space Administration under Grants NNX12AG05G and NNX14AD33G issued through the Astrophysics Research and Analysis (APRA) program (PI: R. Soummer) and by Jet Propulsion Laboratory subcontract No.1539872 (Segmented-Aperture Coronagraph Design and Analysis; PI: R. Soummer).

It is also partly funded by the French Aerospace Lab (ONERA) in the frame of the VASCO Research Project and by the Laboratoire d'Astrophysique de Marseille (LAM).

The authors would also like to thank Gregory Brady for his valuable comments on this paper.

\appendix
\section{The impact of the linear term on the contrast}
\label{a_linear}

In the expression~\ref{eq:13}, the contrast is a sum of a constant term (the deep contrast, intrinsic to the coronagraph) and a quadratic term. However, in the equation~\ref{eq:10}, the intensity, whom the contrast derives from, is a sum of a constant term, a linear term, and a quadratic term. Why can the linear term not be taken into account, while the constant term can?

The linear term is called $L(\mathbf{u})$.
\begin{equation}
\begin{split}
    L(\mathbf{u}) & = 2 \Re(\widehat{P}^*(\mathbf{u}) i \widehat{Z_l}(\mathbf{u}) \sum_{k=1}^{n_{seg}} a_{k,l} e^{-i \mathbf{r_k} . \mathbf{u}}) \\
     & = 2 \Re(\sum_{k=1}^{n_{seg}} a_{k,l} \widehat{P}^*(\mathbf{u}) i \widehat{Z_l}(\mathbf{u}) e^{-i \mathbf{r_k} . \mathbf{u}}) \\
     & = 2 \sum_{k=1}^{n_{seg}} a_{k,l} \Re(\widehat{P}^*(\mathbf{u}) i \widehat{Z_l}(\mathbf{u}) e^{-i \mathbf{r_k} . \mathbf{u}})
\end{split}
\end{equation}
since the coefficients $a_{k,l}$ are real.

- $P$ is real and symmetrical, so its Fourier Transform $\widehat{P}$ is also real and symmetrical, and $\widehat{P}^*$ is too.

- Furthermore, we have:
\begin{equation}
\begin{split}
    i e^{-i \mathbf{r_k} . \mathbf{u}} & = i \cos(-\mathbf{r_k} . \mathbf{u}) + i^2 \sin(-\mathbf{r_k} . \mathbf{u}) \\
     & = i \cos(\mathbf{r_k} . \mathbf{u}) + \sin(\mathbf{r_k} . \mathbf{u})
\end{split}
\end{equation}

- $Z_l$ is necessary real, but can be symmetrical or anti-symmetrical. We study separately these two cases:

If $Z_l$ is real and symmetrical, then $\widehat{Z_l}$ is real and symmetrical. Then:
\begin{equation}
    \Re(\widehat{P}(\mathbf{u}) i \widehat{Z_l}(\mathbf{u}) e^{-i \mathbf{r_k} . \mathbf{u}}) = \widehat{P}(\mathbf{u}) \widehat{Z_l}(\mathbf{u}) \sin(\mathbf{r_k} . \mathbf{u})
\end{equation}
In this case, since $\widehat{P}(\mathbf{u})$ and $\widehat{Z_l}(\mathbf{u})$ are symmetrical and $\sin(\mathbf{r_k} . \mathbf{u})$ anti-symmetrical, then $L$ is anti-symmetrical. As a result, the contrast resulting from computing its average on a symmetrical dark hole is null. 

If $Z_l$ is a real and anti-symmetrical, then $\widehat{Z_l}$ is imaginary and anti-symmetrical. Then:
\begin{equation}
    \Re(\widehat{P}(\mathbf{u}) \widehat{Z_l}(\mathbf{u}) e^{-i \mathbf{r_k} . \mathbf{u}}) = - \widehat{P}(\mathbf{u}) \Im(\widehat{Z_l}(\mathbf{u})) \cos(\mathbf{r_k} . \mathbf{u})
\end{equation}
$\widehat{P}(\mathbf{u})$ and $\cos(\mathbf{r_k} . \mathbf{u})$ are symmetrical, $\Im(\widehat{Z_l}(\mathbf{u}))$ is anti-symmetrical, so $L$ is once again anti-symmetrical and its mean contrast on a symmetrical dark hole is null.

As a conclusion, in both cases, the linear term of the initial equation provides a null contrast.

\bibliography{bib}

\begin{thebibliography}{10}

\bibitem{Stark2016}
C.~C. {Stark}, E.~J. {Cady}, M.~{Clampin}, S.~{Domagal-Goldman}, D.~{Lisman},
  A.~M. {Mandell}, M.~W. {McElwain}, A.~{Roberge}, T.~D. {Robinson},
  D.~{Savransky}, S.~B. {Shaklan}, and K.~R. {Stapelfeldt}, ``{A direct
  comparison of exoEarth yields for starshades and coronagraphs},'' in {\em
  Space Telescopes and Instrumentation 2016: Optical, Infrared, and Millimeter
  Wave},  {\em \procspie} {\bf 9904}, p.~99041U, July 2016.

\bibitem{Ruane2017}
G.~{Ruane}, D.~{Mawet}, J.~{Kastner}, T.~{Meshkat}, M.~{Bottom},
  B.~{Femen{\'{\i}}a Castell{\'a}}, O.~{Absil}, C.~{Gomez Gonzalez}, E.~{Huby},
  Z.~{Zhu}, R.~{Jenson-Clem}, {\'E}.~{Choquet}, and E.~{Serabyn}, ``{Deep
  Imaging Search for Planets Forming in the TW Hya Protoplanetary Disk with the
  Keck/NIRC2 Vortex Coronagraph},'' {\em \aj}~{\bf 154}, p.~73, Aug. 2017.

\bibitem{Mawet2017}
D.~{Mawet}, {\'E}.~{Choquet}, O.~{Absil}, E.~{Huby}, M.~{Bottom}, E.~{Serabyn},
  B.~{Femenia}, J.~{Lebreton}, K.~{Matthews}, C.~A. {Gomez Gonzalez},
  O.~{Wertz}, B.~{Carlomagno}, V.~{Christiaens}, D.~{Defr{\`e}re},
  C.~{Delacroix}, P.~{Forsberg}, S.~{Habraken}, A.~{Jolivet}, M.~{Karlsson},
  J.~{Milli}, C.~{Pinte}, P.~{Piron}, M.~{Reggiani}, J.~{Surdej}, and
  E.~{Vargas Catalan}, ``{Characterization of the Inner Disk around HD 141569 A
  from Keck/NIRC2 L-Band Vortex Coronagraphy},'' {\em \aj}~{\bf 153}, p.~44,
  Jan. 2017.

\bibitem{Mawet2016}
D.~{Mawet}, P.~{Wizinowich}, R.~{Dekany}, M.~{Chun}, D.~{Hall}, S.~{Cetre},
  O.~{Guyon}, J.~K. {Wallace}, B.~{Bowler}, M.~{Liu}, G.~{Ruane}, E.~{Serabyn},
  R.~{Bartos}, J.~{Wang}, G.~{Vasisht}, M.~{Fitzgerald}, A.~{Skemer},
  M.~{Ireland}, J.~{Fucik}, J.~{Fortney}, I.~{Crossfield}, R.~{Hu}, and
  B.~{Benneke}, ``{Keck Planet Imager and Characterizer: concept and phased
  implementation},'' in {\em Adaptive Optics Systems V},  {\em \procspie} {\bf
  9909}, p.~99090D, July 2016.

\bibitem{Boccaletti2015}
A.~{Boccaletti}, P.-O. {Lagage}, P.~{Baudoz}, C.~{Beichman}, P.~{Bouchet},
  C.~{Cavarroc}, D.~{Dubreuil}, A.~{Glasse}, A.~M. {Glauser}, D.~C. {Hines},
  C.-P. {Lajoie}, J.~{Lebreton}, M.~D. {Perrin}, L.~{Pueyo}, J.~M. {Reess},
  G.~H. {Rieke}, S.~{Ronayette}, D.~{Rouan}, R.~{Soummer}, and G.~S. {Wright},
  ``{The Mid-Infrared Instrument for the James Webb Space Telescope, V:
  Predicted Performance of the MIRI Coronagraphs},'' {\em \pasp}~{\bf 127},
  p.~633, July 2015.

\bibitem{Krist2007}
J.~E. {Krist}, C.~A. {Beichman}, J.~T. {Trauger}, M.~J. {Rieke},
  S.~{Somerstein}, J.~J. {Green}, S.~D. {Horner}, J.~A. {Stansberry}, F.~{Shi},
  M.~R. {Meyer}, K.~R. {Stapelfeldt}, and T.~L. {Roellig}, ``{Hunting planets
  and observing disks with the JWST NIRCam coronagraph},'' in {\em Techniques
  and Instrumentation for Detection of Exoplanets III},  {\em \procspie} {\bf
  6693}, p.~66930H, Sept. 2007.

\bibitem{Carlotti2011}
A.~{Carlotti}, R.~{Vanderbei}, and N.~J. {Kasdin}, ``{Optimal pupil
  apodizations of arbitrary apertures for high-contrast imaging},'' {\em Optics
  Express}~{\bf 19}, p.~26796, Dec. 2011.

\bibitem{Mazoyer2017}
J.~{Mazoyer}, L.~{Pueyo}, M.~{N'Diaye}, K.~{Fogarty}, L.~{Leboulleux},
  S.~{Egron}, and C.~{Norman}, ``{Capabilities of ACAD-OSM, an active method
  for the correction of aperture discontinuities},'' in {\em Society of
  Photo-Optical Instrumentation Engineers (SPIE) Conference Series},   {\bf
  10400}, p.~104000G, Sept. 2017.

\bibitem{Zimmerman2016}
N.~T. {Zimmerman}, M.~{N'Diaye}, K.~E. {St.~Laurent}, R.~{Soummer}, L.~{Pueyo},
  C.~C. {Stark}, A.~{Sivaramakrishnan}, M.~{Perrin}, R.~J. {Vanderbei}, N.~J.
  {Kasdin}, S.~{Shaklan}, and A.~{Carlotti}, ``{Lyot coronagraph design study
  for large, segmented space telescope apertures},'' in {\em Space Telescopes
  and Instrumentation 2016: Optical, Infrared, and Millimeter Wave},  {\em
  \procspie} {\bf 9904}, p.~99041Y, July 2016.

\bibitem{Pueyo2013}
L.~{Pueyo} and C.~{Norman}, ``{High-contrast Imaging with an Arbitrary
  Aperture: Active Compensation of Aperture Discontinuities},'' {\em \apj}~{\bf
  769}, p.~102, June 2013.

\bibitem{Guyon2014}
O.~{Guyon}, P.~M. {Hinz}, E.~{Cady}, R.~{Belikov}, and F.~{Martinache}, ``{High
  Performance Lyot and PIAA Coronagraphy for Arbitrarily Shaped Telescope
  Apertures},'' {\em \apj}~{\bf 780}, p.~171, Jan. 2014.

\bibitem{Krist2016}
J.~{Krist}, B.~{Nemati}, and B.~{Mennesson}, ``{Numerical modeling of the
  proposed WFIRST-AFTA coronagraphs and their predicted performances},'' {\em
  Journal of Astronomical Telescopes, Instruments, and Systems}~{\bf 2},
  p.~011003, Jan. 2016.

\bibitem{Shi2016}
F.~{Shi}, K.~{Balasubramanian}, R.~{Bartos}, R.~{Hein}, R.~{Lam}, M.~{Mandic},
  D.~{Moore}, J.~{Moore}, K.~{Patterson}, I.~{Poberezhskiy}, J.~{Shields},
  E.~{Sidick}, H.~{Tang}, T.~{Truong}, J.~K. {Wallace}, X.~{Wang}, and D.~W.
  {Wilson}, ``{Low order wavefront sensing and control for WFIRST
  coronagraph},'' in {\em Space Telescopes and Instrumentation 2016: Optical,
  Infrared, and Millimeter Wave},  {\em \procspie} {\bf 9904}, p.~990418, July
  2016.

\bibitem{Lightsey2010}
P.~A. {Lightsey}, D.~{Chaney}, B.~{Gallagher}, B.~{Brown}, K.~{Smith},
  J.~{Lewis}, A.~{Barto}, S.~{Knight}, S.~{Acton}, C.~{Stewart}, and
  N.~{Siegel}, ``{Optical performance for the actively controlled James Webb
  Space Telescope},'' in {\em Space Telescopes and Instrumentation 2010:
  Optical, Infrared, and Millimeter Wave},  {\em \procspie} {\bf 7731},
  p.~77310B, July 2010.

\bibitem{Lightsey2014}
P.~A. {Lightsey}, J.~S. {Knight}, and G.~{Golnik}, ``{Status of the optical
  performance for the James Webb Space Telescope},'' in {\em Space Telescopes
  and Instrumentation 2014: Optical, Infrared, and Millimeter Wave},  {\em
  \procspie} {\bf 9143}, p.~914304, Aug. 2014.

\bibitem{Macintosh2006}
B.~{Macintosh}, M.~{Troy}, R.~{Doyon}, J.~{Graham}, K.~{Baker}, B.~{Bauman},
  C.~{Marois}, D.~{Palmer}, D.~{Phillion}, L.~{Poyneer}, I.~{Crossfield},
  P.~{Dumont}, B.~M. {Levine}, M.~{Shao}, G.~{Serabyn}, C.~{Shelton},
  G.~{Vasisht}, J.~K. {Wallace}, J.-F. {Lavigne}, P.~{Valee}, N.~{Rowlands},
  K.~{Tam}, and D.~{Hackett}, ``{Extreme adaptive optics for the Thirty Meter
  Telescope},'' in {\em Society of Photo-Optical Instrumentation Engineers
  (SPIE) Conference Series},  {\em \procspie} {\bf 6272}, p.~62720N, June 2006.

\bibitem{Kasper2008}
M.~E. {Kasper}, J.-L. {Beuzit}, C.~{Verinaud}, N.~{Yaitskova}, P.~{Baudoz},
  A.~{Boccaletti}, R.~G. {Gratton}, N.~{Hubin}, F.~{Kerber}, R.~{Roelfsema},
  H.~M. {Schmid}, N.~A. {Thatte}, K.~{Dohlen}, M.~{Feldt}, L.~{Venema}, and
  S.~{Wolf}, ``{EPICS: the exoplanet imager for the E-ELT},'' in {\em Adaptive
  Optics Systems},  {\em \procspie} {\bf 7015}, p.~70151S, July 2008.

\bibitem{Davies2010}
R.~{Davies}, N.~{Ageorges}, L.~{Barl}, L.~R. {Bedin}, R.~{Bender},
  P.~{Bernardi}, F.~{Chapron}, Y.~{Clenet}, A.~{Deep}, E.~{Deul}, M.~{Drost},
  F.~{Eisenhauer}, R.~{Falomo}, G.~{Fiorentino}, N.~M. {F{\"o}rster Schreiber},
  E.~{Gendron}, R.~{Genzel}, D.~{Gratadour}, L.~{Greggio}, F.~{Grupp},
  E.~{Held}, T.~{Herbst}, H.-J. {Hess}, Z.~{Hubert}, K.~{Jahnke}, K.~{Kuijken},
  D.~{Lutz}, D.~{Magrin}, B.~{Muschielok}, R.~{Navarro}, E.~{Noyola},
  T.~{Paumard}, G.~{Piotto}, R.~{Ragazzoni}, A.~{Renzini}, G.~{Rousset}, H.-W.
  {Rix}, R.~{Saglia}, L.~{Tacconi}, M.~{Thiel}, E.~{Tolstoy}, S.~{Trippe},
  N.~{Tromp}, E.~A. {Valentijn}, G.~{Verdoes Kleijn}, and M.~{Wegner},
  ``{MICADO: the E-ELT adaptive optics imaging camera},'' in {\em Ground-based
  and Airborne Instrumentation for Astronomy III},  {\em \procspie} {\bf 7735},
  p.~77352A, July 2010.

\bibitem{Quanz2015}
S.~P. {Quanz}, I.~{Crossfield}, M.~R. {Meyer}, E.~{Schmalzl}, and J.~{Held},
  ``{Direct detection of exoplanets in the 3-10 {$\mu$}m range with
  E-ELT/METIS},'' {\em International Journal of Astrobiology}~{\bf 14},
  pp.~279--289, Apr. 2015.

\bibitem{Yaitskova2003}
N.~{Yaitskova}, K.~{Dohlen}, and P.~{Dierickx}, ``{Analytical study of
  diffraction effects in extremely large segmented telescopes},'' {\em Journal
  of the Optical Society of America A}~{\bf 20}, pp.~1563--1575, Aug. 2003.

\bibitem{Trauger2007}
J.~T. {Trauger} and W.~A. {Traub}, ``{A laboratory demonstration of the
  capability to image an Earth-like extrasolar planet},'' {\em \nat}~{\bf 446},
  pp.~771--773, Apr. 2007.

\bibitem{Baudoz2012}
P.~{Baudoz}, J.~{Mazoyer}, M.~{Mas}, R.~{Galicher}, and G.~{Rousset}, ``{Dark
  hole and planet detection: laboratory results using the self-coherent
  camera},'' in {\em Ground-based and Airborne Instrumentation for Astronomy
  IV},  {\em \procspie} {\bf 8446}, p.~84468C, Sept. 2012.

\bibitem{Mazoyer2014}
J.~{Mazoyer}, P.~{Baudoz}, R.~{Galicher}, and G.~{Rousset}, ``{High-contrast
  imaging in polychromatic light with the self-coherent camera},'' {\em
  \aap}~{\bf 564}, p.~L1, Apr. 2014.

\bibitem{Dalcanton2015}
J.~{Dalcanton}, S.~{Seager}, S.~{Aigrain}, S.~{Battel}, N.~{Brandt},
  C.~{Conroy}, L.~{Feinberg}, S.~{Gezari}, O.~{Guyon}, W.~{Harris},
  C.~{Hirata}, J.~{Mather}, M.~{Postman}, D.~{Redding}, D.~{Schiminovich},
  H.~P. {Stahl}, and J.~{Tumlinson}, ``{From Cosmic Birth to Living Earths: The
  Future of UVOIR Space Astronomy},'' {\em ArXiv e-prints} , July 2015.

\bibitem{Pueyo2017}
L.~{Pueyo}, N.~{Zimmerman}, M.~{Bolcar}, T.~{Groff}, C.~{Stark}, G.~{Ruane},
  J.~{Jewell}, R.~{Soummer}, K.~{St.~Laurent}, J.~{Wang}, D.~{Redding},
  J.~{Mazoyer}, K.~{Fogarty}, R.~{Juanola-Parramon}, S.~{Domagal-Goldman},
  A.~{Roberge}, O.~{Guyon}, and A.~{Mandell}, ``{The LUVOIR architecture ``A''
  coronagraph instrument},'' in {\em Society of Photo-Optical Instrumentation
  Engineers (SPIE) Conference Series},  {\em Society of Photo-Optical
  Instrumentation Engineers (SPIE) Conference Series} {\bf 10398}, p.~103980F,
  Sept. 2017.

\bibitem{Mennesson2016}
B.~{Mennesson}, S.~{Gaudi}, S.~{Seager}, K.~{Cahoy}, S.~{Domagal-Goldman},
  L.~{Feinberg}, O.~{Guyon}, J.~{Kasdin}, C.~{Marois}, D.~{Mawet}, M.~{Tamura},
  D.~{Mouillet}, T.~{Prusti}, A.~{Quirrenbach}, T.~{Robinson}, L.~{Rogers},
  P.~{Scowen}, R.~{Somerville}, K.~{Stapelfeldt}, D.~{Stern}, M.~{Still},
  M.~{Turnbull}, J.~{Booth}, A.~{Kiessling}, G.~{Kuan}, and K.~{Warfield},
  ``{The Habitable Exoplanet (HabEx) Imaging Mission: preliminary science
  drivers and technical requirements},'' in {\em Space Telescopes and
  Instrumentation 2016: Optical, Infrared, and Millimeter Wave},  {\em
  \procspie} {\bf 9904}, p.~99040L, July 2016.

\bibitem{Stahl2013}
H.~P. {Stahl}, M.~{Postman}, and W.~S. {Smith}, ``{Engineering specifications
  for large aperture UVO space telescopes derived from science requirements},''
  in {\em UV/Optical/IR Space Telescopes and Instruments: Innovative
  Technologies and Concepts VI},  {\em \procspie} {\bf 8860}, p.~886006, Sept.
  2013.

\bibitem{Stahl2015}
M.~T. {Stahl}, S.~B. {Shaklan}, and H.~P. {Stahl}, ``{Preliminary analysis of
  effect of random segment errors on coronagraph performance},'' in {\em
  Techniques and Instrumentation for Detection of Exoplanets VII},  {\em
  \procspie} {\bf 9605}, p.~96050P, Sept. 2015.

\bibitem{Greenhouse2016}
M.~A. {Greenhouse}, ``{The JWST science instrument payload: mission context and
  status},'' in {\em Space Telescopes and Instrumentation 2016: Optical,
  Infrared, and Millimeter Wave},  {\em \procspie} {\bf 9904}, p.~990406, July
  2016.

\bibitem{Clampin2008}
M.~{Clampin}, ``{Status of the James Webb Space Telescope (JWST)},'' in {\em
  Space Telescopes and Instrumentation 2008: Optical, Infrared, and
  Millimeter},  {\em \procspie} {\bf 7010}, p.~70100L, July 2008.

\bibitem{Borde2006}
P.~J. {Bord{\'e}} and W.~A. {Traub}, ``{High-Contrast Imaging from Space:
  Speckle Nulling in a Low-Aberration Regime},'' {\em \apj}~{\bf 638},
  pp.~488--498, Feb. 2006.

\bibitem{Malbet1995}
F.~{Malbet}, J.~W. {Yu}, and M.~{Shao}, ``{High-Dynamic-Range Imaging Using a
  Deformable Mirror for Space Coronography},'' {\em \pasp}~{\bf 107}, p.~386,
  Apr. 1995.

\bibitem{Quirrenbach2005}
A.~{Quirrenbach}, ``{Coronographic Methods for the Detection of Terrestrial
  Planets},'' {\em ArXiv Astrophysics e-prints} , Feb. 2005.

\bibitem{Cavarroc2006}
C.~{Cavarroc}, A.~{Boccaletti}, P.~{Baudoz}, T.~{Fusco}, and D.~{Rouan},
  ``{Fundamental limitations on Earth-like planet detection with extremely
  large telescopes},'' {\em \aap}~{\bf 447}, pp.~397--403, Feb. 2006.

\bibitem{Janinpotiron2018}
P.~{Janin-Potiron}, P.~{Martinez}, and M.~{Carbillet}, ``{Analytical
  decomposition of Zernike and hexagonal modes over an hexagonal segmented
  optical aperture},'' {\em Applied Optics (submitted)} , 2018.

\bibitem{Mahajan2006}
V.~N. {Mahajan} and G.-M. {Dai}, ``{Orthonormal polynomials for hexagonal
  pupils},'' {\em Optics Letters}~{\bf 31}, pp.~2462--2464, Aug. 2006.

\bibitem{janinpotiron2017}
P.~Janin-Potiron, {\em {Active correction of pupil discontinuities on segmented
  telescopes for high contrast imaging and high angular resolution}}.
\newblock Theses, {Universit{\'e} C{\^o}te d'Azur}, Oct. 2017.

\bibitem{BornWolf1999}
M.~{Born} and E.~{Wolf}, {\em {Principles of optics : electromagnetic theory of
  propagation, interference and diffraction of light}}, 1999.

\bibitem{Tuthill2000}
P.~G. {Tuthill}, J.~D. {Monnier}, W.~C. {Danchi}, E.~H. {Wishnow}, and C.~A.
  {Haniff}, ``{Michelson Interferometry with the Keck I Telescope},'' {\em
  \pasp}~{\bf 112}, pp.~555--565, Apr. 2000.

\bibitem{Lacour2011}
S.~{Lacour}, P.~{Tuthill}, P.~{Amico}, M.~{Ireland}, D.~{Ehrenreich},
  N.~{Huelamo}, and A.-M. {Lagrange}, ``{Sparse aperture masking at the VLT. I.
  Faint companion detection limits for the two debris disk stars HD 92945 and
  HD 141569},'' {\em \aap}~{\bf 532}, p.~A72, Aug. 2011.

\bibitem{Postman2012}
M.~{Postman}, T.~{Brown}, K.~{Sembach}, M.~{Giavalisco}, W.~{Traub},
  K.~{Stapelfeldt}, D.~{Calzetti}, W.~{Oegerle}, R.~{Michael Rich}, H.~{Phillip
  Stahl}, J.~{Tumlinson}, M.~{Mountain}, R.~{Soummer}, and T.~{Hyde},
  ``{Advanced Technology Large-Aperture Space Telescope: science drivers and
  technology developments},'' {\em Optical Engineering}~{\bf 51},
  pp.~011007--011007--12, Jan. 2012.

\bibitem{Feinberg2014}
L.~D. {Feinberg}, A.~{Jones}, G.~{Mosier}, N.~{Rioux}, D.~{Redding}, and
  M.~{Kienlen}, ``{A cost-effective and serviceable ATLAST 9.2m telescope
  architecture},'' in {\em Space Telescopes and Instrumentation 2014: Optical,
  Infrared, and Millimeter Wave},  {\em \procspie} {\bf 9143}, p.~914316, Aug.
  2014.

\bibitem{Paul2014}
B.~{Paul}, J.-F. {Sauvage}, L.~M. {Mugnier}, K.~{Dohlen}, T.~{Fusco}, and
  M.~{Ferrari}, ``{Simultaneous phase and amplitude retrieval with COFFEE: from
  theory to laboratory results},'' in {\em Ground-based and Airborne
  Instrumentation for Astronomy V},  {\em \procspie} {\bf 9147}, p.~91479O,
  July 2014.

\bibitem{Soummer2003}
R.~{Soummer}, C.~{Aime}, and P.~E. {Falloon}, ``{Stellar coronagraphy with
  prolate apodized circular apertures},'' {\em \aap}~{\bf 397}, pp.~1161--1172,
  Jan. 2003.

\bibitem{NDiaye2015a}
M.~{N'Diaye}, L.~{Pueyo}, and R.~{Soummer}, ``{Apodized Pupil Lyot Coronagraphs
  for Arbitrary Apertures. IV. Reduced Inner Working Angle and Increased
  Robustness to Low-order Aberrations},'' {\em \apj}~{\bf 799}, p.~225, Feb.
  2015.

\bibitem{NDiaye2016ApJ}
M.~{N'Diaye}, R.~{Soummer}, L.~{Pueyo}, A.~{Carlotti}, C.~C. {Stark}, and M.~D.
  {Perrin}, ``{Apodized Pupil Lyot Coronagraphs for Arbitrary Apertures. V.
  Hybrid Shaped Pupil Designs for Imaging Earth-like planets with Future Space
  Observatories},'' {\em \apj}~{\bf 818}, p.~163, Feb. 2016.

\bibitem{Stahl2016b}
H.~P. {Stahl}, ``{Advanced Mirror Technology Development (AMTD) project:
  overview and year four accomplishments},'' in {\em Advances in Optical and
  Mechanical Technologies for Telescopes and Instrumentation II},  {\em
  \procspie} {\bf 9912}, p.~99120S, July 2016.

\bibitem{Stahl2016c}
M.~T. {Stahl}, H.~P. {Stahl}, and S.~B. {Shaklan}, ``{Contrast Leakage as a
  function of telescope motion},'' {\em Mirror Technology Days} , 2016.

\end{thebibliography}
\bibliographystyle{spiebib}

\end{document}